\documentclass[times,twocolumn,final]{elsarticle}

\usepackage{medima}
\usepackage{framed,multirow}

\usepackage{url}
\usepackage[table]{xcolor}
\usepackage{graphicx}
\usepackage{amsmath}
\usepackage{amsthm}
\usepackage{amsfonts}
\usepackage{amssymb}
\usepackage{booktabs}
\usepackage[hidelinks]{hyperref}
\usepackage{bm} 
\usepackage{tabularx} 
\usepackage{booktabs} 
\usepackage{makecell} 
\usepackage{mathtools}
\usepackage{multirow}
\usepackage{cleveref}
\usepackage{subcaption}

\newcommand{\expnumber}[2]{{#1}\mathrm{e}{#2}}
\renewcommand{\vec}[1]{\text{#1}}
\newcommand{\mat}[1]{\text{#1}}
\newcommand{\interval}[1]{{\textpm#1}}
\newcommand{\pred}[1]{{#1}^\mathrm{p}}
\newcommand{\gt}[1]{{#1}^\mathrm{gt}}
\newcommand{\Mpred}{\pred{\mathcal{M}}}
\newcommand{\Mgt}{\gt{\mathcal{M}}}
\newcommand{\Ppred}{\pred{\mathcal{Q}}}
\newcommand{\Pgt}{\gt{\mathcal{Q}}}
\newcommand{\stimes}{{\times}} 
\newcommand{\rev}{\color{black}}
\newcommand{\revv}{\color{black}}

\DeclareMathOperator*{\argmin}{arg\,min}
\DeclarePairedDelimiter\abs{\lvert}{\rvert}%
\DeclarePairedDelimiter\norm{\lVert}{\rVert}%

\definecolor{newcolor}{rgb}{.8,.349,.1}

\journal{Medical Image Analysis}

\begin{document}

\verso{Fabian Bongratz \textit{et~al.}}

\begin{frontmatter}

\title{Neural deformation fields for template-based reconstruction of cortical surfaces from MRI}%

\author[1,3]{Fabian \snm{Bongratz}\corref{cor1}}
\cortext[cor1]{Corresponding author. 
\textit{E-mail:} fabi.bongratz@tum.de
}
\author[1,2]{Anne-Marie \snm{Rickmann}}
\author[1,2,3]{Christian \snm{Wachinger}}

\address[1]{Laboratory for Artificial Intelligence in Medical Imaging, Technical University of Munich, Munich 81675, Germany}
\address[2]{Department of Child and Adolescent Psychiatry, Ludwig-Maximilians-University, Munich 80336, Germany}
\address[3]{Munich Center for Machine Learning, Munich, Germany}

\begin{abstract}
The reconstruction of cortical surfaces is a prerequisite for quantitative analyses of the cerebral cortex in magnetic resonance imaging (MRI). 
Existing segmentation-based methods separate the surface registration from the surface extraction, which is computationally inefficient and prone to distortions.
We introduce Vox2Cortex-Flow (V2C-Flow), a deep mesh-deformation technique that learns a deformation field from a brain template to the cortical surfaces of an MRI scan. To this end, we present a geometric neural network that models the deformation-describing ordinary differential equation in a continuous manner. The network architecture comprises convolutional and graph-convolutional layers, which allows it to work with images and meshes at the same time.
V2C-Flow is not only very fast, requiring less than two seconds to infer all four cortical surfaces, but also establishes vertex-wise correspondences to the template during reconstruction.
In addition, V2C-Flow is the first approach for cortex reconstruction that models white matter and pial surfaces jointly, therefore avoiding intersections between them. 
Our comprehensive experiments on internal and external test data demonstrate that V2C-Flow results in cortical surfaces that are state-of-the-art in terms of accuracy. Moreover, we show that the established correspondences are more consistent than in FreeSurfer 
and that they can directly be utilized for cortex parcellation and group analyses of cortical thickness.  
\end{abstract}


\end{frontmatter}

\section{Introduction}

\subsection{Motivation}

The cerebral cortex is a thin and tightly folded sheet of neural tissue that is closely related to brain functionality~\citep{fjellwalhovd_brain_changes}. 
In vivo studies with magnetic resonance imaging (MRI) have identified cortical changes associated with aging, diseases, and neuropsychology~\citep{SHAW2016age_thinning,lerch2005,kuperberg2003,YOTTER2011fractal_dimension}. 
Identifying such cortical alterations in MRI is remarkable, given that they are in the sub-millimeter range and the common resolution of brain MRI scans is 1mm. 
Computational tools, like FreeSurfer~\citep{fischl1999cortical,fischl2012freesurfer}, have been instrumental for such quantitative analyses of the cortex as they provide highly accurate reconstructions using geometric surfaces. 
The estimation of these surfaces, i.e., triangular meshes, enables a reconstruction accuracy that goes below the voxel resolution~\citep{fischl2012freesurfer}.

\begin{figure*}
    \centering
    \includegraphics[width=0.8\textwidth]{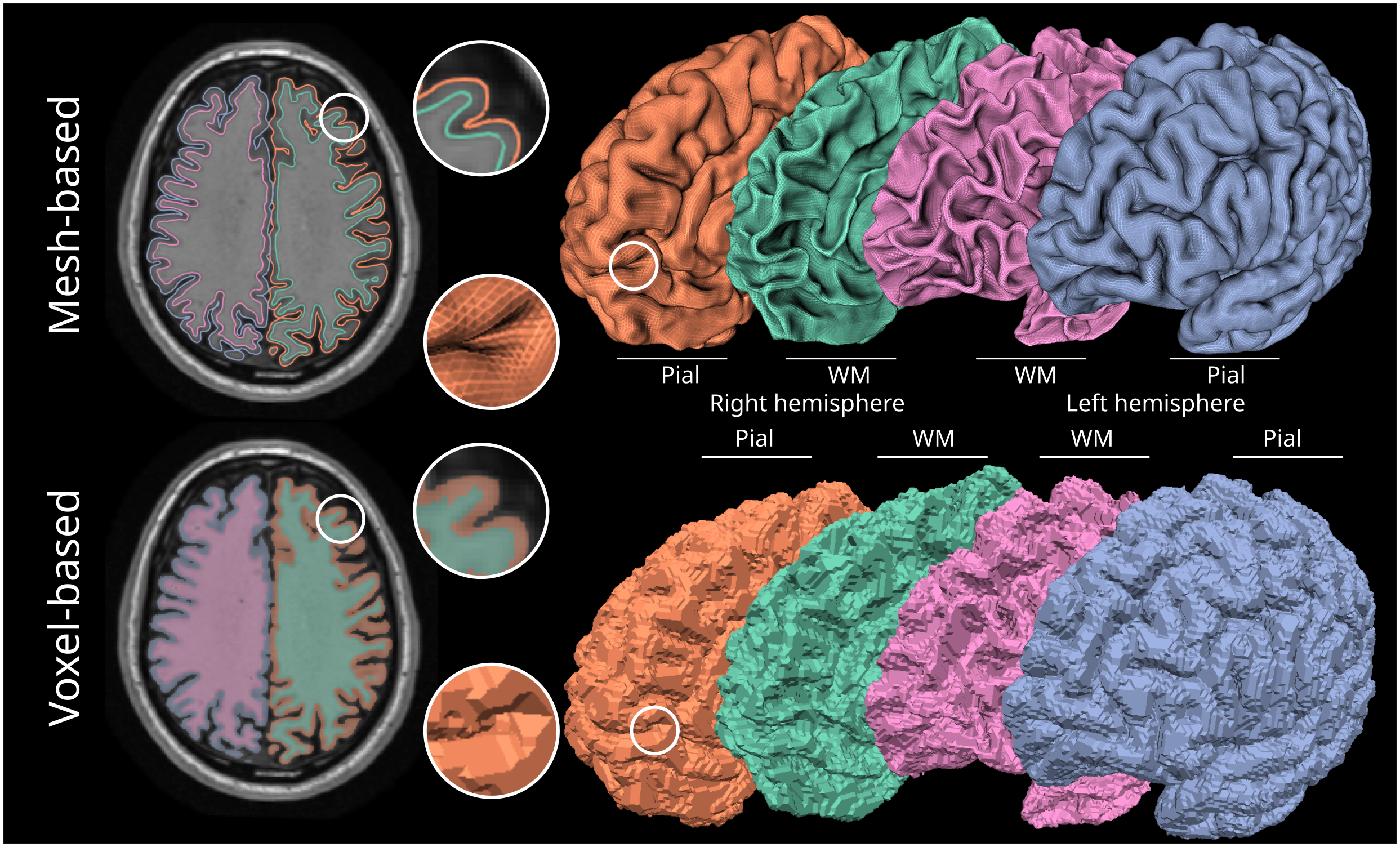}
    \caption{The smooth and tightly folded geometry of the cortex is better represented by mesh-based surface extraction methods than by voxel-based segmentation methods. Inner white matter (WM) and outer pial surfaces of each hemisphere were extracted from a T1-weighted MRI scan with the proposed V2C-Flow method (mesh-based) and the popular nnUNet~\citep{isensee2020} with subsequent application of Marching Cubes (voxel-based). While voxel-based methods attempt to classify each voxel, mesh-based approaches focus on delineating the contours between distinct tissue types; this fundamental difference is depicted on the left for an axial slice of the MRI scan.}
    \label{fig:voxelvsmesh}
\end{figure*}

The cortex reconstruction in FreeSurfer independently estimates white matter (WM) and pial surfaces for each subject (i.e., the interface between gray and white matter, and the interface between gray matter and cerebrospinal fluid).
For follow-up analyses, e.g., cortical parcellation or group comparisons, subject meshes are registered to a template
to create point-wise correspondences~\citep{Fischl1999HighresolutionIA,yeo2010registration}. 
Due to the complex geometry of the cortex, this involves a sophisticated pipeline with inflating the cortex, mapping it to the sphere, and a spherical registration. 
This process inevitably creates distortions through the spherical mapping~\citep{fischl1999cortical} and is time-consuming.

Deep learning-based methods have recently led to tremendous advances in whole-brain segmentation and can reduce the processing time to seconds~\citep{roy2019,moeskops2016brainsegmentation,Wachinger2018,HUO2019105,roy2022are,henschel2020}. 
However, these methods operate in voxel space and are not well suited for the accurate reconstruction of the intricate cortex geometry and do not support the computation of commonly used biomarkers like curvature or surface area. 
Even though a surface can be extracted from the voxel segmentation via algorithms like Marching Cubes~\citep{lorensen1987,lewiner2003efficient}, the obtained results are usually not satisfactory due to partial volume effects and staircase artifacts. These artifacts can be well-observed in the bottom row of \Cref{fig:voxelvsmesh} and they confound further analyses.
Another learning-based approach relies on deep signed distance functions (SDFs), which provide an alternative implicit surface representation~\citep{mescheder2019,park2019deepsdf,sitzmann2020implicit,xu2019}.
Even though deep SDFs were successfully applied for cortical surface reconstruction~\citep{santacruz2021}, the required topology correction and mesh extraction is computationally intensive and prone to introducing geometric artifacts into the extracted surfaces~\citep{Bongratz_2022_CVPR,lebrat2021corticalflow}. Furthermore, correspondences between surfaces can only be established during post-processing in these methods.

\subsection{Contributions}

To address the mentioned shortcomings, we present a novel approach for cortical surface reconstruction termed Vox2Cortex-Flow (or briefly V2C-Flow). 
In contrast to voxel- and SDF-based approaches, V2C-Flow explicitly models the cortical surfaces with triangular meshes. This results in smooth and accurate three-dimensional representations of the cortex, see \Cref{fig:voxelvsmesh}.
The underlying idea is to take a template mesh as an input to the network and to learn a vertex-wise deformation field conditioned on 3D brain scans. 
For a new input scan, V2C-Flow starts from template cortex meshes and applies a sequence of deformations to reconstruct the cortex.
The main challenge here is to generate smooth and regular output meshes, which we 
achieve by numerical integration of a deformation-describing ordinary differential equation (ODE) and adequate geometric regularization based on local curvature. 
V2C-Flow directly models the ODE as a neural function on the mesh, while prior work only worked on single points~\citep{gupta2020} and in the image domain~\citep{lebrat2021corticalflow}. 

By doing so, V2C-Flow pushes the current technical boundaries, as the four cortex meshes have a total of 655,000 vertices that are deformed simultaneously. 
The large number of vertices is needed to achieve a high reconstruction accuracy of the tightly folded cortex. 
This is in large contrast to applications in computer vision about multi-view  reconstruction~\citep{wang2018pixel2mesh,wen2019}, where networks for explicit surface reconstruction typically work with less than 10,000 vertices per surface. 
The training of V2C-Flow was made possible by efficient implementation and the latest generation of graphics processing units (GPUs) that provide large enough memory for learning from high-resolution meshes. 
For the training and validation of V2C-Flow, we worked with {\rev six} different publicly available brain datasets to demonstrate the generalization across datasets. 
The trained V2C-Flow models, together with the respective training and inference code, are publicly available so that the results can be reproduced and the model deployed on new data\footnote{\url{https://github.com/ai-med/Vox2Cortex}}.

The distinguishing characteristics of V2C-Flow are: 
\begin{itemize}
    \item The reconstruction of the cortex via deformation of the template creates inherent correspondences between subject and template meshes. 
    This permits to directly propagate a template parcellation to the subject or to directly perform group analyses on a per-vertex basis, thereby eliminating the need for computationally intensive and error-prone surface inflation and registration.
    \item V2C-Flow computes all four cortical surfaces in 1.6 seconds, thanks to the parallel execution on GPUs and efficient implementation. This is about 10K times faster than FreeSurfer, opening up new avenues for clinical translation and immensely facilitating the processing of large-scale population studies.  
    \item V2C-Flow simultaneously estimates all four surfaces, which allows to learn and impose relationships between them. This is in particular relevant for white matter and pial surfaces of the same hemisphere since they are the boundaries of the gray matter and hence should not intersect.
    \item V2C-Flow can be readily integrated into existing neuroimaging pipelines as it can deal with various kinds of brain templates, e.g., FreeSurfer's fsaverage or age-specific population templates. The usage of fsaverage enables re-using the rich tools in FreeSurfer for cortex analysis. 
\end{itemize}

This article substantially extends our previous work on cortical surface reconstruction~\citep{Bongratz_2022_CVPR} by introducing, for the first time, numerically integrable graph deformation blocks (graph NODEs). This reduces the number of network parameters without losing any expressiveness. 
Moreover, we improve the results and include a large variety of new experiments.
More precisely, we (i) initialize the surface reconstruction with FreeSurfer's fsaverage and population-specific reference templates, which allows us to (ii) solve cortex parcellation as an additional task at minimal extra cost. Further, we (iii) assess the geometric consistency of reconstructed points learned with the curvature-weighted Chamfer loss, which challenges the common approach of separating the cortex segmentation from the registration. Finally, we (iv) include {\rev four} more recent baselines and (v) report results on {\rev three} new datasets with varying scanner types, demographics, diagnoses{\rev, and manually set landmark points.}

\section{Related work}
In the following, we first discuss traditional neuroimaging frameworks capable of extracting cortical surface meshes from MRI. Afterward, we review more recent learning-based approaches for image-conditioned surface reconstruction, which can be distinguished into implicit and explicit methods according to the primarily used surface representation. 

\textit{Traditional neuroimaging frameworks.}
Established software for the analysis of brain MRI scans~\citep{dale1999cortical,fischl1999cortical,smith2004advances,shattuck2002} relies on traditional (non-deep learning) image processing pipelines. These pipelines usually include skull stripping, intensity normalization, voxel-based tissue classification, topology correction, mesh extraction, spherical inflation, and registration. While these frameworks are known to be very reliable since they often comprise several decades of algorithmic development, they suffer from lengthy computation times and cause unfavorable delays in cortex measurements. In particular, traditional brain analysis tools ignore recent advances in general-purpose computation on GPUs and deep learning, which has the potential to drastically speed up the brain analysis pipeline. Yet, traditional neuroimaging frameworks typically offer lots of functionality for downstream analyses, e.g., for surface parcellation, group analyses, or visualization, which is more difficult to implement with recent approaches that do not come with such inherent functionality.

\textit{Implicit surface reconstruction.}
Learning-based implicit surface reconstruction methods represent a two-dimensional surface $\mathcal{S} \subset \mathbb{R}^3$ implicitly via a function $f: \mathbb{R}^3 \rightarrow \mathbb{R}$, respectively its zero-level set, i.e., $\mathcal{S} = \{p \in \mathbb{R}^3 | f(p) = 0\}$. For the extraction of practically relevant surface meshes, these methods usually rely on a mesh-extraction algorithm like marching cubes~\citep{lorensen1987, lewiner2003efficient}. In the image domain, $f$ can either represent a discrete signed distance function (SDF), or it may indicate the occupancy of voxels to a certain tissue class (similar to a segmentation map). However, voxel-based surface representations are usually not accurate enough to represent the tight folds of the cortex. Hence, voxel-based methods have been paired with subsequent mesh extraction and deformation~\citep{henschel2020,ma2022_cortexODE} or smoothing~\citep{gopinath2021} to delineate tissue boundaries with sub-voxel accuracy. 
To alleviate the discrete nature of voxel-based representations, deep neural networks can be leveraged to parameterize continuous SDFs, yielding so-called deep SDFs. Deep SDFs have become popular in the computer vision domain~\citep{mescheder2019,park2019deepsdf,sitzmann2020implicit,xu2019} due to their expressive power and their ability to produce surfaces of arbitrary resolution. Recently, they have also been applied for the task of cortical surface reconstruction~\citep{santacruz2021}.
Although the versatility of SDFs is advantageous for the reconstruction of shapes with unknown characteristics, ensuring the spherical topology of cortical sheets requires the application of an intricate topology-correction routine. This post-processing is not just inconvenient, it is also prone to introduce geometric artifacts into the extracted surfaces~\citep{Bongratz_2022_CVPR,lebrat2021corticalflow}. 
Further, meshes extracted with Marching Cubes from an SDF do not come with inter-subject correspondences, which makes it difficult to compare them on a per-vertex basis.

\textit{Mesh-based surface reconstruction.}
Instead of relying on an implicit representation of tissue boundaries, mesh-based methods~\citep{groueix2018,wang2019,KONG2021deep,ma2021b,wang2018pixel2mesh,wen2019,wickramasinghe2020} directly parameterize the surfaces explicitly via a (usually triangular) mesh $\mathcal{M}=\{\mathcal{V}, \mathcal{F}\}$, consisting of vertices $\mathcal{V}$ and faces $\mathcal{F}$. These methods take a template mesh as input and learn to deform it to the target shape conditioned on 2D~\citep{groueix2018,wang2019,wang2018pixel2mesh,wen2019} or 3D input images~\citep{KONG2021deep,wickramasinghe2020,ma2021b,lebrat2021corticalflow,Bongratz_2022_CVPR}. This bears the potential to create inter- and intra-subject shape correspondences~\citep{groueix2018_3dcoded}, which, however, has neither been explored nor validated yet for the complex geometry of the human cortex. In addition, the mesh deformation enables us to engrave the sphere-like surface topology directly into the template mesh, thereby preventing holes and bridges. Still, the challenge is to compute a smooth, ideally diffeomorphic deformation to obtain watertight output meshes. {\rev In practice, however, such an optimal solution can only be approximated} by loss regularization~\citep{wang2018pixel2mesh,wickramasinghe2020} or through numerical integration of ODEs~\citep{gupta2020,lebrat2021corticalflow}. 
For the first time, we use a graph neural network (GNN) to model the deformation-describing ODE in a continuous manner and train it with a locally re-weighted ratio of point and regularization loss terms --- hence combining the best out of both worlds.
Probably most related to our work are CorticalFlow($^{++}$)~\citep{lebrat2021corticalflow,santacruz2022_corticalflow++} and TopoFit~\citep{hoopes2022topofit}. Both of these methods start with a generic brain-shaped template and deform it to the cortical boundaries represented in the input MRI. However, CorticalFlow($^{++}$) consists of multiple UNet-shaped networks that are trained in an iterative manner, which is intricate and can take up to several weeks. From an architectural perspective, TopoFit is similar to Vox2Cortex~\citep{Bongratz_2022_CVPR} as it is made up of a combination of convolutional and graph-convolutional neural networks. However, TopoFit only supports the reconstruction of white matter surfaces while we aim for a holistic method for the entire cortex to be able to model existing dependencies among cortical boundaries. 

\section{Methods}
\begin{figure*}
    \centering
    \includegraphics[width=\textwidth]{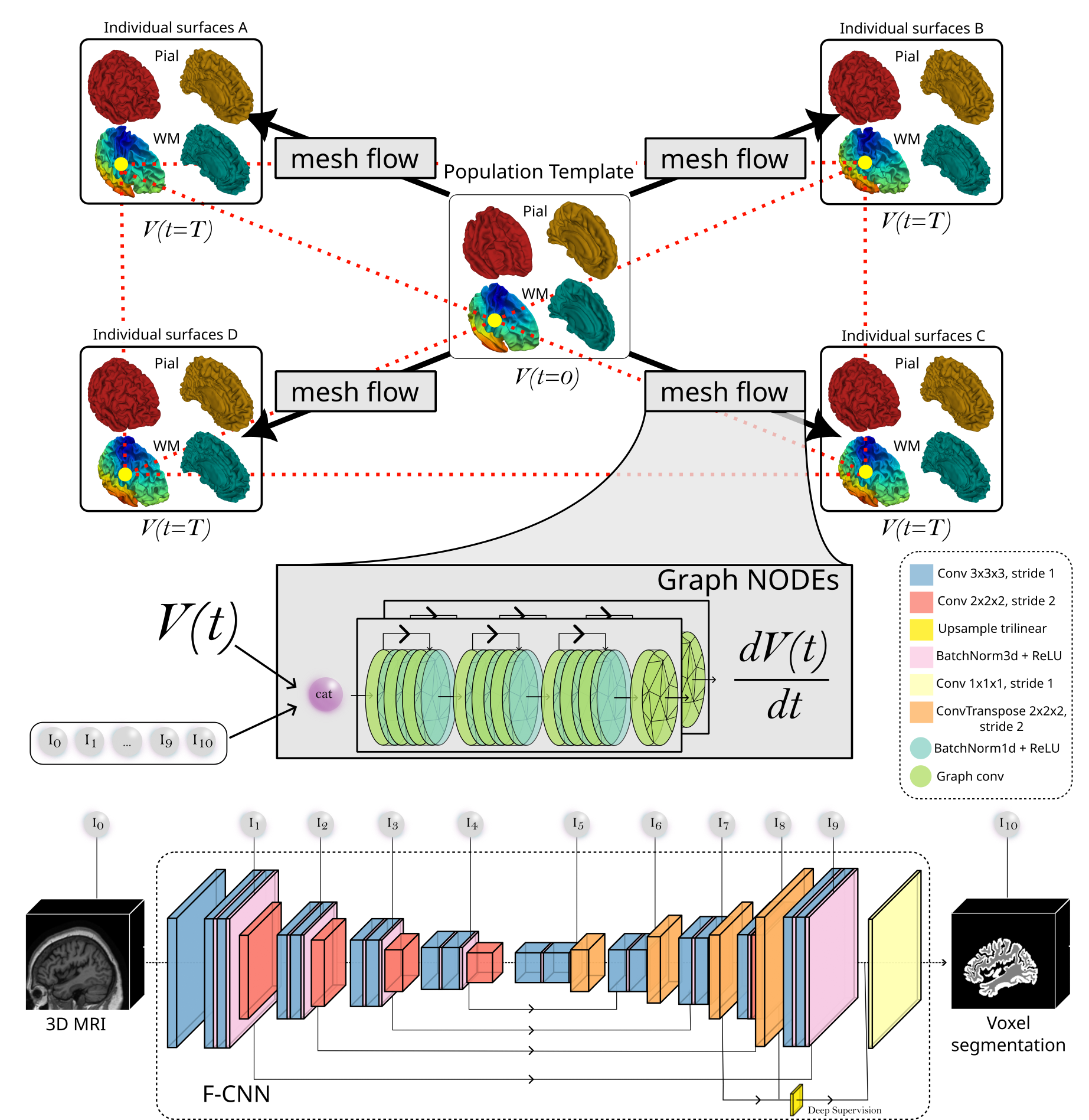}
    \caption{V2C-Flow extracts inner white matter (WM) and outer (pial) cortical surface meshes from MRI. Computed surfaces have point-wise correspondences to the input template. The rainbow-colored surfaces are real distinct surfaces extracted by V2C-Flow, with corresponding vertices having identical colors. To this end, a deformation field relative to the template's vertices $V(t)$, modeled as a neural ordinary differential equation (NODE), is predicted. The deformation is conditioned on the input MRI scan. The boundary condition of the ODE is given at $t=0$ by the template. The network consists of a fully-convolutional neural network (F-CNN) and a graph neural network (GNN), operating in image and mesh space, respectively. Both networks are connected by a feature-sampling module that maps image features onto vertices. Skip connections within CNN blocks are omitted for the sake of clarity.
    }
    \label{fig:architecture}
\end{figure*}

Vox2Cortex-Flow, or briefly V2C-Flow, is a deep learning-based method for cortical surface reconstruction from MRI, which maintains point-wise correspondences to an input template.
\Cref{fig:architecture} depicts an architectural overview of V2C-Flow, which takes a 3D brain MRI scan together with a brain-shaped mesh template as input. 
The output comprises the four cortical surfaces, i.e., inner WM and outer pial boundaries of each hemisphere. As a side product, a voxel-based segmentation of gray and white matter is also produced. 
The surfaces are the result of applying a flow field onto the template vertices, keeping the connectivity of the mesh fixed during its evolution. Thereby, V2C-Flow establishes point-wise correspondences between vertices of individual brain reconstructions to the input template and thus across subjects. 
We illustrate these correspondences in \Cref{fig:architecture} by emblematical connections between the template and four subjects (A, \ldots, D) and by assigning an exclusive color to each vertex.

\subsection{Graph NODEs}
In this work, we deal with two-dimensional manifolds $\mathcal{M} \subset \mathbb{R}^3$, which are embedded in 3D Euclidean space and represented as triangular meshes $\mathcal{M}=\{\mathcal{V}, \mathcal{F}\}$. Each mesh consists of $\abs{\mathcal{V}}$ vertices, represented as a matrix $V \in \mathbb{R}^{\abs{\mathcal{V}}\times3}$, and $\abs{\mathcal{F}}$ faces, represented by $F\in \mathbb{\rev N}^{\abs{\mathcal{F}}\times3}$ that stores the indices of the respective vertices for each triangle. Since we desire to keep the sphere-like topology of the input meshes in the output surfaces, the surface reconstruction boils down to computing a displacement $f: \mathbb{R}^{n\times3} \rightarrow \mathbb{R}^{n\times3}$ for the vertices $V$. 

Encouraged by recent work in the field of diffeomorphic shape transformations~\citep{gupta2020,lebrat2021corticalflow,ma2022_cortexODE}, we model the deformation of the brain template to the target shapes with $S \in \mathbb{N}$ autonomous ordinary differential equations (ODEs),
\begin{equation} \label{eq:ODEsingle}
    \frac{dV(t)}{dt} = f^{\rev (s)}( \mathcal{I}, V(t)), \quad s \in \{0, \ldots, S-1\},
\end{equation}
where $V(t)$ are the coordinates of the vertices of the mesh at time $t \in \left[t_{1,s}, t_{2,s}\right]$ and $\mathcal{I}=\{I_0, \ldots, I_{10}\}$ is a set of latent features extracted from the input scan. Without loss of generality, we assume that $t_{2,s} - t_{1,s} = 1$ for all $ s \in \{0, \ldots, S-1\}$ and $t_{1, 0} = 0$. The entire deformation can formally be described as the time-dependent ODE of the form
\begin{equation} \label{eq:ODE}
    \frac{dV(t)}{dt} = f(t, \mathcal{I}, V(t)) = f^{\rev (\lfloor t \rfloor)}( \mathcal{I},V(t)).
\end{equation}
The boundary condition is given by the vertices of the input template at $t=0$ and the final prediction is obtained from the vertices $V(S)$. To solve the ODE, we apply a forward Euler integration scheme with step size $h$, i.e., $V^{\rev (k+1)} = V^{\rev (k)} + h f^{\rev (\lfloor t_k \rfloor)}( \mathcal{I}, V^{\rev (k)})$. {\rev See Supplementary Figure~2 for a visualization of the deformation process. During the integration, hypercolumn features are re-evaluated at moving vertex locations. {\revv Compared to non-integrable graph deformation blocks~\citep{wickramasinghe2020,hoopes2022topofit,Bongratz_2022_CVPR}, the graph NODEs enable us to obtain better reconstruction accuracy at lower memory cost. More precisely, with the Euler integration scheme, the GPU memory required for inference is independent of the number of integration steps and scales linearly at training time; see Supplementary Figure~3 for a thorough analysis.} While, in theory, it would be possible to apply more involved ODE solvers like Midpoint or Runge-Kutta integration, the current limitations of GPUs prohibited training on large templates with these solvers.} 

Each deformation field $f^{\rev (s)}$ is represented by a graph neural network (GNN) block, comprising three graph-residual blocks. Each graph-residual block encompasses three spectrum-free graph convolutions from~\citep{morris2019} (GraphConv layers), batch normalization layers, and ReLU activations in alternation. One GraphConv layer transforms the input vertex features $\vec{f}_i \in \mathbb{R}^{d_\mathrm{in}}$ of a vertex $\vec{v}_i \in \mathcal{V}$ to the output $\vec{f}'_i \in \mathbb{R}^{d_\mathrm{out}}$ by aggregating
\begin{equation} \label{eq:graph conv}
	\vec{f}'_i =  \frac{1}{1 + \abs{\mathcal{N}(i)}}\left[\mat{W}_0 \vec{f}_i + \mat{W}_1  \sum_{j \in \mathcal{N}(i)} \vec{f}_j   + \vec{b}\right],
\end{equation}
where $\mat{W}_0, \mat{W}_1 \in \mathbb{R}^{d_\mathrm{out} \stimes d_\mathrm{in}}$ and the bias $\vec{b} \in \mathbb{R}^{d_\mathrm{out}}$. $\mathcal{N}(i)$ is the set of neighbors of $\vec{v}_i$ in the template mesh. Since we deal with fixed meshes, we can use the efficient sparse implementation of GraphConvs in~\citep{Fey/Lenssen/2019} (unless we use group-specific templates, e.g., conditioned on age or sex). The vertex-wise input features $\vec{f}$ are given by a concatenation of image features $\mathcal{I}_{0, \ldots, 10}$, similar to hypercolumns~\citep{HariharanAGM15_hypercolumns}, the vertex coordinates $\vec{v}$, and deep vertex features extracted by the previous GNN block.
For the initial GNN block, these features are created from another GNN block that takes the coordinates from the template and a surface ID (0 for WM, 1 for pial surfaces) as input.

\subsection{Hypercolumn extraction}
For the mesh deformation to be meaningful, i.e., to ultimately match the tissue boundaries in the brain scan, it needs to be conditioned on the input image. To this end, we employ a residual 3D UNet architecture~\citep{cicek2016,isensee2020,ronneberger2015,zhang2018} to extract image features $\mathcal{I}_{1, \ldots, 9}$ at multiple resolutions. Together with the input image $\mathcal{I}_{0}$ and the output segmentation $\mathcal{I}_{10}$ they make up the hypercolumns that are mapped onto mesh vertices by trilinear interpolation. \Cref{fig:architecture} depicts the UNet architecture in V2C-Flow. It comprises a total of four down- and up-sampling stages. Each stage consists of residual convolution blocks with batch-normalization layers~\citep{ioffe2015}, ReLU activations~\citep{nair2010}, and (transposed) convolutional layers with stride 2. From the decoder, we divert deep-supervision branches~\citep{zeng2017} to propagate the segmentation loss directly to the lower decoder layers. The output of the CNN is a three-class segmentation map (white matter, gray matter, and background), computed via Softmax activations.
While previous work used hypercolumns either only from the image encoder or decoder~\citep{wang2018pixel2mesh, wickramasinghe2020, KONG2021deep}, experiments in~\citep{Bongratz_2022_CVPR} showed that providing features from the entire UNet is most effective. In this work, we add the input image and the segmentation output to the set of hypercolumn features, making up a total of eleven feature maps (four encoder stages, four decoder stages, the bottleneck, input image, and output segmentation).

\subsection{Interdependence between inner and outer brain surfaces}
White matter and pial surfaces are intrinsically related as they represent the boundaries of the cerebral sheet. 
We model this interdependence in V2C-Flow inherently by deforming all surfaces simultaneously in a single forward pass, i.e., by sharing the deformation field's parameters among the surfaces. Moreover, we introduce \emph{virtual edges} between corresponding vertices of the inner and outer surfaces. {\rev These edges add an additional neighbor to each vertex from the respective other surface, and they are virtual in that they only exist in the mesh that is processed by the GNN but not in the output meshes. Thereby, features from the inner and outer surfaces are aggregated in each graph-convolutional layer.}
This facilitates the alignment of inner and outer cortex surfaces as it renders the information exchange between the respective flow fields possible{\revv; see Supplementary Table~4 for an ablation study}.
Noteworthy, none of the related deep cortical surface reconstruction methods model the interdependence between surfaces and estimate them independently, which can lead to an implausible intersection of white and gray matter surfaces (cf.~\Cref{fig:comparison}). 

\subsection{Mesh template}
In general, mesh template-based surface reconstruction prevents intrinsic topological artifacts like bridges or holes in the surfaces. While, in principle, any mesh template with sphere-like topology can be used as input into our model, starting from a generic brain-shaped template has shown to improve the surface accuracy compared to even more generic shapes like ellipsoids~\citep{Bongratz_2022_CVPR}. In this work, we propose to employ a population template, like FreeSurfer's \emph{fsaverage} subject (163,842 vertices per surface and  $163,842 \times 4 = 655,368$ vertices in total), or age-specific population templates of similar size as starting point for the cortex reconstruction. This enables us to exploit per-vertex correspondences of the predicted surfaces to the template for downstream applications, e.g., atlas-based parcellation or group analyses. The usage of fsaverage has the additional benefit that our surfaces can be compared to pre-existing FreeSurfer meshes.

\subsection{Loss function} \label{sec:loss}
We train V2C-Flow with network weights $\theta$ in an end-to-end fashion by minimizing a combination of voxel and mesh loss, i.e.,
\begin{equation} \label{eq:argmin_model}
    \theta = \underset{\theta}{\argmin} \left[ \mathbb{E}_{(x,\gt{y}) \thicksim \mathcal{D}} \left[ \mathcal{L}_{\text{vox}}(\pred{y}, \gt{y}) + \mathcal{L}_{\text{mesh}}(\pred{y}, \gt{y}) \right] \right],
\end{equation}
where $\pred{y}=\text{V2C-Flow}_\theta(x)$ comprises the four cortical surfaces 
$\{\Mpred_{s,c} \vert s = 0, \ldots, S-1, c=0, \ldots 3\}$ from $S$ graph NODEs and voxel segmentations $\{ \pred{B}_l \vert  l = 0, \ldots, L-1\}$ from the final UNet output and $L-1$ deep supervision branches. $\gt{y} = (\Mgt, \gt{B})$ are corresponding ground-truth surfaces and segmentations for a certain input scan $x\in \mathbb{R}^{H\stimes W\stimes D}$ from the training data $\mathcal{D}$.

\subsubsection{Voxel loss}
We use the standard cross-entropy loss for the voxel segmentation task of our model. Let $\mathcal{L}_{\text{CE}}(\pred{B}_l, \gt{B})$ be the cross-entropy loss between a predicted segmentation map $\pred{B}_l \in [0, 1]^{K\stimes H\stimes W\stimes D}$ and a label $\gt{B} \in \{0, 1\}^{K \stimes H\stimes W\stimes D}$ with $K$ classes. The voxel loss of V2C-Flow is 
\begin{equation} \label{eq:voxel loss}
	\mathcal{L}_{\text{vox}}(\pred{y}, \gt{y}) = \sum_{l=1}^{L} \mathcal{L}_\text{CE}(\pred{B}_l, \gt{B}).
\end{equation}

\subsubsection{Mesh loss}
We leverage the curvature-weighted Chamfer distance $\mathcal{L}_\text{CwC}$ introduced in~\citep{Bongratz_2022_CVPR} to assure geometric accuracy. In addition, we train with two regularizing terms: the edge loss $\mathcal{L}_\text{edge}$ and the (intra-mesh) normal consistency loss $\mathcal{L}_\text{NC}$. Intuitively, the regularizers enforce the regularity and smoothness of the output surfaces. In summary, the mesh loss is defined as
\begin{equation}
\begin{split}
    \mathcal{L}_\text{mesh}(\pred{y}, \gt{y}) 
    &= \sum_{s,c} \left[ \mathcal{L}_\text{CwC} (\Mpred_{s,c}, \Mgt_c) \right. \\
    &+ \left. \lambda_1 \mathcal{L}_\text{edge}(\Mpred_{s,c}) + \lambda_2 \mathcal{L}_\text{NC}(\Mpred_{s,c}) \right].  
\end{split}
\end{equation}
From a log-spaced grid search, we found $\lambda_1=1.0$ and $\lambda_2=0.0001$ to work best for cortical surfaces. 

\textit{Curvature-weighted Chamfer loss.}
A major challenge in deformation-based surface reconstruction methods is to set the ratio between accuracy and regularization terms in the loss function. This is, in particular, challenging in the brain as the regularization could easily lead to a deficit in accuracy in highly curved regions (cf.~Supplementary Figure~1). To address this issue, we leverage the curvature-weighted Chamfer loss~\citep{Bongratz_2022_CVPR}, which re-weights the ratio between point and regularization losses locally based on ground-truth curvature. It gives higher importance to the point loss in highly curved regions and it is defined as
\begin{equation} \label{eq:weighted_Chamfer distance}
	\begin{split}	
	\mathcal{L}_\text{CwC} (\Mpred, \Mgt) 
	&= \frac{1}{\abs{\Pgt}} \sum_{\vec{u} \in \Pgt} \kappa(\vec{u}) \underset{\vec{v} \in \Ppred}{\min} \norm{\vec{u} - \vec{v}}^2 \\
	&+ \frac{1}{\abs{\Ppred}} \sum_{\vec{v} \in \Ppred} \kappa(\tilde{\vec{u}}) \underset{\vec{u} \in \Pgt}{\min} \norm{\vec{v} - \vec{u}}^2,
	\end{split}
\end{equation}
where $\tilde{\vec{u}} =\argmin_{\vec{r} \in \Pgt} \norm{\vec{v} - \vec{r}}^2$. $\Ppred=\{q_i \mid q_i \in \Mpred \wedge 1\leq i \leq N\}$ and $\Pgt=\{q_i \mid q_i \in \Mgt \wedge 1\leq i \leq N\}$ are $N=50,000$ predicted and ground-truth points, respectively, sampled from the surfaces in a differentiable manner~\citep{gkioxari2019,smith2019}. $\kappa(\vec{p}) \in \left[1, \kappa_{\max}\right]$ is a point weight based on the local discrete mean curvature~\citep{meyer2003}. We set $\kappa_{\max}=5$ as in~\citep{Bongratz_2022_CVPR}.

\textit{Edge loss.}
The edge loss ensures the regularity of predicted edges and it is defined as
\begin{equation}
	\mathcal{L}_\text{edge}(\mathcal{M}) = \frac{1}{\abs{\mathcal{E}}} \sum_{(i,j) \in \mathcal{E}} \norm{\vec{v}_i - \vec{v}_j}^2,
\end{equation}
where $\mathcal{E}$ is the set of edges of the mesh.

\textit{Normal consistency loss.}
The normal consistency loss enforces the smoothness of the deformed mesh by penalizing differences in the orientation of adjacent faces. It is defined as
\begin{equation}
	\mathcal{L}_\text{NC}(\mathcal{M}) = \frac{1}{\abs{\mathcal{E}}} \sum_{\substack{f_0 \cap f_1 \in \mathcal{E}  \\ f_0, f_1 \in \mathcal{F}}} 1 - \cos(\vec{n}(f_0), \vec{n}(f_1)),
\end{equation}
where $\vec{n}(f)$ assigns a normal to each {\rev face $f$} of the mesh.

\section{Experimental setting}
\subsection{Datasets and preprocessing} \label{sec:datasets}
All data used in this work was processed using FreeSurfer v7.2~\citep{fischl2012freesurfer} to extract white matter and pial surfaces. 
MRI scans (orig.mgz files) were registered to the MNI152 space using rigid and subsequent affine registration. When used as supervised training labels, FreeSurfer meshes were simplified to about 40,000 vertices per surface using quadric edge collapse decimation~\citep{garland1997} implemented in MeshLab~\citep{cignoniMeshLab2008}. {\rev This reduces the memory footprint for training and we did not observe any benefit from training with larger surfaces. For the evaluation, however, we employ the original, not decimated FreeSurfer meshes.} We zero-padded all input images to have shape $192 \stimes 208 \stimes 192$. Intensity values were min-max-normalized into $[0, 1]$.

To demonstrate the performance of V2C-Flow, we use the public datasets described in the following. A tabular overview can be found in Supplementary Table~1.

\textit{OASIS.}
The OASIS-1 dataset~\citep{oasis} contains MRI T1 scans of 416 subjects. 100 subjects have been diagnosed with very mild to moderate Alzheimer's disease. We split the data balanced on diagnosis, age, and sex, resulting in 292, 44, and 80 subjects for training, validation, and testing, respectively.

\textit{ADNI.}
The Alzheimer's Disease Neuroimaging Initiative (ADNI, \url{http://adni.loni.usc.edu}) provides MRI T1 scans for subjects with Alzheimer's Disease, Mild Cognitive Impairment, and healthy subjects.
After removing data with processing artifacts (e.g., the scan shown in Supplementary Figure~6), we split the data into training, validation, and testing set, balanced according to diagnosis, age, and sex. As ADNI is a longitudinal study, we only use the initial (baseline) scan for each subject. 
In our experiments, we use a subset of the ADNI data which contains 1,155 subjects for training, 169 for validation, and 323 for testing.

\textit{Mindboggle.}
The Mindboggle-101 dataset~\citep{kleinMindboggle2012} contains MRI T1 scans of 101 subjects, from different data sources. We have removed one scan (Colin-27) from the data as it had severe artifacts. When training on Mindboggle, we split it randomly into 75/5/20 scans for training/validation/test. Further, the Mindboggle data can be distinguished according to the study in which it has originally been acquired (12 from HLN, 23 from MMRR, 22 from NKI-RS, 20 from NKI-TRT, 20 from OASIS-TRT, and 3 from other individuals).  

\textit{J-ADNI.}
The Japanese ADNI Project (J-ADNI) (\url{https://www.j-adni.org/}) provides MRI T1 scans from healthy subjects and subjects with Alzheimer's disease from Japan. We use 502 baseline scans, of which 352/49/101 are used for training/validation/testing, again balanced according to age, sex, and diagnosis.

\textit{Test-retest.}
The public test-retest (TRT) dataset from~\citep{maclaren2014} contains 120 MRI T1 scans from three subjects, where each subject has been scanned twice in 20 days. We use the dataset to assess the consistency of reconstructed points and surfaces (testing only).

\textit{\rev Manual landmarks (JHU).}
{\rev The dataset from~\citep{Shiee2013_jhu_data} contains five MRI T1 scans from healthy subjects and five subjects diagnosed with multiple sclerosis (MS). For every scan, two raters placed 840 landmarks each to delineate seven cortical subregions characterized by certain sulci or gyri. An additional 100 landmarks per MS subject are provided from three raters, respectively, near WM lesions in MS patients. In our experiments, we do not train on this data and use it for the evaluation of trained models only.}

\subsection{Implementation and parameters}

We implemented V2C-Flow based on PyTorch (v1.10.0)~\citep{paszke2019pytorch}, torch-geometric (v2.0.4)~\citep{Fey/Lenssen/2019},
and PyTorch3d (v0.6.1)~\citep{ravi2020}. 
We trained our models on two (four for ADNI) Nvidia A100 GPUs with 80GB memory, mixed precision, and PyTorch's distributed data-parallel framework. On each GPU node, we used a batch size of one. For inference, we leveraged an Nvidia A6000 GPU.
We optimize V2C-Flow with AdamW~\citep{loshchilov2018decoupled} ($\beta_1 = 0.9$, $\beta_2 = 0.999$), a cyclic learning rate scheduler~\citep{Smith2017CyclicalLR} (base learning rates $\expnumber{1}{-4}$ and $\expnumber{5}{-5}$ for the CNN and the GNN, respectively), and we train our models until convergence on the validation set with a hard limit of 105 epochs (corresponds to about three days of training). We found the initialization of output GraphConv layers with zero weights to be crucial for training success. All other initial GNN weights are sampled from a normal distribution (mean=0, SD=0.01). Our UNet has channel sizes of $(16, 32, 64, 128, 256, 64, 32, 16, 8)$ and a GNN block has $64$ channels per GraphConv layer. UNet weights are sampled initially from the standard uniform distribution in PyTorch. Per default, we employ $S=2$ graph NODEs and $L=3$ segmentation outputs. 
Inference of a trained V2C-Flow model takes about $1.6$ seconds on the A6000 GPU and requires 2.4GB of VRAM (with two graph NODEs, and five Euler integration steps, see Supplementary Figure~3 for an analysis of these parameters). Compared to Vox2Cortex~\citep{Bongratz_2022_CVPR}, this corresponds to a reduction of about 30\% in memory (2.4GB vs.~3.4GB) and 90\% in  inference time (1.6s vs.~18s). 

\section{Results and discussion}\label{sec:results}

\subsection{Surface accuracy and inference time} \label{sec:accuracy}

\begin{table*}[htb]
\renewcommand\bfdefault{b}
\small
\centering
\setlength{\tabcolsep}{3.5pt} 
\caption{Comparison of V2C-Flow with state-of-the-art cortex reconstruction methods on the ADNI test set. We report the average symmetric surface distance (ASSD) and the 90-percentile Hausdorff distance (HD$_{90}$) in mm (lower is better). Values are averaged over the test set and we denote standard deviations over the test set with \interval{X}. Best and second-best values are \textbf{highlighted}. For V2C-Flow, (P) indicates an age-specific population template and S/F the smooth/folded fsaverage template. {\rev $^*$} TopoFit has been adapted for the reconstruction of pial surfaces as well.}
\begin{tabular}{lllllllll}
    \toprule
      Method  &   \multicolumn{2}{c}{Left WM surface} 
    & \multicolumn{2}{c}{Right WM surface}
     & \multicolumn{2}{c}{Left pial surface} 
    & \multicolumn{2}{c}{Right pial surface} 
      \\ \midrule

    & \multicolumn{1}{c}{ASSD} 
    & \multicolumn{1}{c}{HD$_{90}$}  
    & \multicolumn{1}{c}{ASSD} 
    & \multicolumn{1}{c}{HD$_{90}$} 
    & \multicolumn{1}{c}{ASSD} 
    & \multicolumn{1}{c}{HD$_{90}$}  
    & \multicolumn{1}{c}{ASSD} 
    & \multicolumn{1}{c}{HD$_{90}$} 


    \\
    \cmidrule(lr){2-2} \cmidrule(lr){3-3} \cmidrule(lr){4-4} \cmidrule(lr){5-5} \cmidrule(lr){6-6} \cmidrule(lr){7-7} \cmidrule(lr){8-8} \cmidrule(lr){9-9}

    V2C-Flow (S) & 
    .179 \interval{.041} & 
    \textbf{.393 \interval{.095}} & 
    .177 \interval{.031} & 
    \textbf{.389 \interval{.076}}  &
    \textbf{.176 \interval{.030}} & 
    .400 \interval{.066} & 
    \textbf{.174 \interval{.022}} & 
    \textbf{.389 \interval{.055}} 
    \\
    
    V2C-Flow (P) & 
    \textbf{.176 \interval{.041}} & 
    .398 \interval{.098} & 
    \textbf{.174 \interval{.030}} & 
    .393 \interval{.077} &
    .177 \interval{.029} & 
    .405 \interval{.065} & 
    .176 \interval{.022} & 
    .402 \interval{.053} 
    \\ 

    V2C-Flow (F) & 
    .183 \interval{.040} & 
    .407 \interval{.097} & 
    .182 \interval{.032} & 
    .405 \interval{.081} &
    .181 \interval{.030} & 
    .415 \interval{.067} & 
    .179 \interval{.024} & 
    .406 \interval{.057} 
   \\

    \addlinespace[.3em]

    V2C~\citep{Bongratz_2022_CVPR} &
    .197 \interval{.041} & 
    .435 \interval{.100} & 
    .198 \interval{.031} & 
    .431 \interval{.076} &
    .210 \interval{.033} & 
    .500 \interval{.094} & 
    .216 \interval{.028} & 
    .515 \interval{.084}
    \\

   CF~\citep{lebrat2021corticalflow} & 
   .209 \interval{.040} & 
   .479 \interval{.101} & 
   .208 \interval{.031} & 
   .478 \interval{.085} &
   .216 \interval{.033} & 
   .519 \interval{.072} & 
   .215 \interval{.024} & 
   .516 \interval{.062} 
   \\

   CF$^{++}$~\citep{santacruz2022_corticalflow++}  &
   .181 \interval{.038} &
   .401 \interval{.089}&
   .181 \interval{.032}&
   .401 \interval{.080} &
   \textbf{.169 \interval{.034}}& 
   \textbf{.375 \interval{.069}} &
   \textbf{.169 \interval{.028}} &
   \textbf{.375 \interval{.059}}
   \\

    DeepCSR~\citep{santacruz2021}  & 
    .422 \interval{.058} &
    .852 \interval{.134} & 
    .420 \interval{.058}  & 
    .880 \interval{.156} &
    .454 \interval{.059} & 
    .927 \interval{.243} & 
    .422 \interval{.053}& 
    .890 \interval{.197}
    \\

    \rev CODE~\citep{ma2022_cortexODE} &
    \rev \textbf{.172 \interval{.044}} & 
    \rev \textbf{.367 \interval{.096}} & 
    \rev \textbf{.172 \interval{.034}} & 
    \rev \textbf{.363 \interval{.077}} & 
    \rev .183 \interval{.035} & 
    \rev \textbf{.381 \interval{.070}} & 
    \rev .191 \interval{.034} & 
    \rev .393 \interval{.066}   
    \\

    \rev TopoFit~\citep{hoopes2022topofit} &
    \rev .211 \interval{.039} & 
    \rev .469 \interval{.096} & 
   \rev .210 \interval{.032} & 
   \rev .477 \interval{.083} & 
   \rev .217 \interval{.036}$^*$ & 
   \rev .490 \interval{.084}$^*$ & 
   \rev .247 \interval{.034}$^*$ & 
   \rev .557 \interval{.081}$^*$ \\
     
    \\

    \bottomrule
\end{tabular}

\label{tab:comparison}
    
\end{table*}

An important aspect of cortical surfaces is the obtained accuracy since cortical folds are often tighter than the typical scan resolution of 1mm per voxel. 
We evaluate the accuracy of V2C-Flow and compare it to recent state-of-the-art cortex reconstruction methods using the ADNI data. Quantitatively, we evaluate the accuracy of all methods with respect to the FreeSurfer~\citep{fischl2012freesurfer} (v7.2) silver standard in \Cref{tab:comparison}. We compute the average symmetric surface distance (ASSD) and 90-percentile Hausdorff distance (HD$_{90}$) from 100,000 randomly sampled surface points (see Supplementary Material for details).

In the comparison, we include V2C-Flow models trained with three different templates. Namely, we use (i) the standard folded fsaverage template (F), (ii) an extensively smoothed fsaverage template (S) {\rev obtained after application of 50 steps of Laplacian smoothing~\citep{vollmer1999}, which is similar to the templates used in~\citep{Bongratz_2022_CVPR,lebrat2021corticalflow,santacruz2022_corticalflow++}}, and (iii) a population template generated from the ADNI training set in age brackets of ten years (P) to have a tailored template.

We observe in \Cref{tab:comparison} that V2C-Flow yields an average (ASSD, $<0.185$mm) and worst-case (HD$_{90}$, $<0.420$mm) deviation from FreeSurfer that is far below the voxel resolution of 1mm for all input templates. {\rev We achieve the best results mostly (not without exception and not by a large margin) with the smoothed fsaverage template. This is surprising as it indicates that the population-specific geometric information contained in the folded templates is of limited value for the learned reconstruction. Yet, working with population templates allows for direct mapping of supplementary vertex information, e.g., an atlas, to the output surfaces through the established correspondences, cf.~\Cref{sec:parcellation}. Such information is typically not available for custom templates, for instance, the one proposed in \citep{santacruz2022_corticalflow++}. Moreover, the independence of the used template is an important property since it offers the opportunity to integrate V2C-Flow into current neuroimaging pipelines by applying existing templates.}

\begin{figure*}
    \centering
    \includegraphics[width=\textwidth]{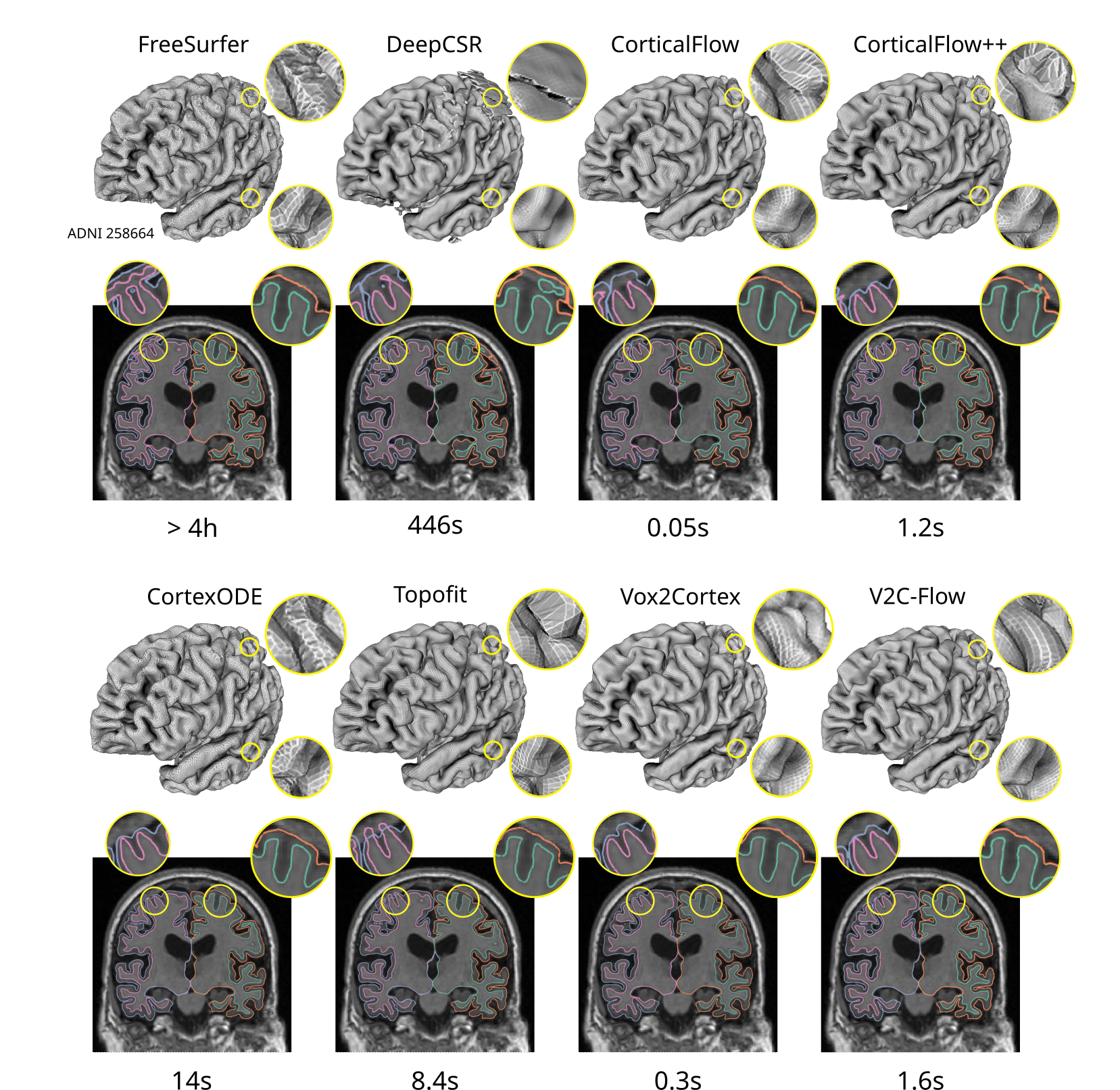}
    \vspace*{1mm}
    \caption{\rev V2C-Flow yields accurate and smooth surfaces while alternative methods introduce artifacts in this sample from the ADNI test set. The required inference time is measured in seconds (s) or hours (h) for the reconstruction of the four cortical surfaces of one subject on an Nvidia A6000 GPU for DL-based methods and on a single CPU for FreeSurfer {\revv (v7.2)}.}
    \label{fig:comparison}
\end{figure*}

In comparison to other DL-based methods for cortical surface reconstruction, V2C-Flow performs comparable to the best alternative method{\rev s}, i.e., CorticalFlow$^{++}$~(CF$^{++}$) {\rev and CortexODE~(CODE), with V2C-Flow being in the forefront on both, WM and pial, surfaces.} We attribute the highest accuracy of CF$^{++}$, {\rev CODE,} and V2C-Flow to the continuous-time deformation process that they have in common. In terms of training time, though, V2C-Flow is about 10 times faster than CF$^{++}$, taking three days instead of four weeks. The long training time in CF$^{++}$ is probably due to the compute-intensive ODE-solver (Runge-Kutta method with 30 integration steps) and the separate modeling of the deformation fields (three UNets per surface). {\rev CODE, on the other hand, is slower in terms of inference time (14s vs.~1.6s in V2C-Flow, cf.~\cref{fig:comparison}) due to the sequential segmentation and mesh extraction approach and does not come with point correspondences to a template.} 
Other DL-based baseline methods, either template-based (Vox2Cortex~\citep{Bongratz_2022_CVPR} (V2C), CorticalFlow~\citep{lebrat2021corticalflow} (CF), {\rev TopoFit~\citep{hoopes2022topofit}}) or SDF-based (DeepCSR~\citep{santacruz2021}), are clearly outperformed in terms of surface accuracy. 
{\rev Supported by the observations from \Cref{fig:comparison} and Supplementary Table~2, we further argue that the separate modeling of surfaces in CF/CF$^{++}$ and TopoFit bears the risk of creating many anatomically implausible intersections between WM and pial surfaces. Instead, the joint modeling in V2C-Flow reduces the number of intersecting WM and pial surfaces. Further artifacts of existing methods like intersections with the skull (CF, CF$^{++}$) and distorted surfaces (DeepCSR, FreeSurfer) are also alleviated in V2C-Flow as shown in \Cref{fig:comparison}. 
}

In terms of inference time, all deep learning-based methods offer a drastic speedup from several hours to seconds compared to traditional CPU-based frameworks like FreeSurfer. This is mainly due to the parallel processing on GPUs. Yet, the topology correction and surface extraction routine in DeepCSR {\rev and CortexODE requires a couple of seconds or minutes, depending on the implementation}. Template-based methods ({\rev TopoFit}, CF, CF$^{++}$, V2C, V2C-Flow) are the fastest as they do not require topology correction and surface extraction; they directly yield usable cortex meshes.

{\rev
\subsection{Robustness to white matter lesions}

\begin{table}[!ht]
\renewcommand\bfdefault{b}
    \centering
    \small
    \setlength{\tabcolsep}{2.7pt} 
    \caption{\rev Distance of manual landmarks to reconstructed white matter (WM) and pial surfaces in the vicinity of  WM lesions in patients diagnosed with multiple sclerosis. Values indicate the mean distance of the landmarks to the mesh surface (\interval{standard deviation}) in mm for three different raters. The best overall value is \textbf{highlighted}.}
    {\rev
    \begin{tabular}{lcccccl}
    
    \toprule
    
         & Rater & CF$^{++}$ & CODE  & TopoFit & V2C-Flow (S) & \multicolumn{1}{c}{FS72} \\ \midrule

         \multirow{3}{*}{pial} & A & 
        .57 \interval{.37} & 
        .80 \interval{.45} & 
        .70 \interval{.41} 	& 
        .47 \interval{.40} &
        .43 \interval{.41}  
\\ 
        
         & B & 
         .46 \interval{.39} & 
         .60 \interval{.42} & 
         .53 \interval{.38} & 
         .45 \interval{.60} &
         .46 \interval{.64}  

         \\
         
         & C & 
         .76 \interval{.44} & 
         .93 \interval{.47} & 
         .84 \interval{.45} 	& 
         .66 \interval{.58} &
         .67 \interval{.64} 

         \\ 

          \multirow{3}{*}{WM} & A & 
          .44 \interval{.35} & 
          .38 \interval{.42} & 
          .49 \interval{.46}& 
          .48 \interval{.32} &
          .62 \interval{1.0} 

          \\ 
          
         & B & 
         .55 \interval{.56} & 
         .67 \interval{.67} & 
         .65 \interval{.63} 	& 
         .56 \interval{.53} &
         .77 \interval{1.1}  

         \\ 
         
         & C & 
         .64 \interval{.54} & 
         .73 \interval{.76} & 
         .78 \interval{.79} & 
         .67 \interval{.56} &
         .84 \interval{.87} 

         \\

         \midrule

         \multicolumn{2}{c}{all} &
         .57 \interval{.46} &
         .68 \interval{.58} &
         .66 \interval{.55} &
        \textbf{.55 \interval{.52}} &
         .63 \interval{.83} 

        \\

        \bottomrule
    \end{tabular}
    }
    \label{tab:accuracy_lesion}
\end{table}

Although the evaluation of surface accuracy with respect to FreeSurfer has become a standard practice~\citep{santacruz2021,lebrat2021corticalflow,santacruz2022_corticalflow++,hoopes2022topofit,Bongratz_2022_CVPR,ma2022_cortexODE}, it can be problematic in difficult regions where FreeSurfer itself might be inaccurate. For instance, white matter lesions complicate the cortex reconstruction, especially if they lie near the cortical boundaries~\cite{Shiee2013_jhu_data}. To assess the robustness of V2C-Flow to such pathological changes, we perform an evaluation with respect to manual landmarks using data from five subjects diagnosed with multiple sclerosis (MS)~\citep{Shiee2013_jhu_data}. We evaluate the accuracy in the vicinity of WM lesions in \Cref{tab:accuracy_lesion} and observe that, on average over the three raters and both surfaces, V2C-Flow surpasses all methods including FreeSurfer. See Supplementary Figure~7 for an exemplary visualization of landmarks and predicted contour next to a WM lesion. In Supplementary Table~3, we report further results on the rest of the manual annotations in the JHU dataset segregated by diagnosis, rater, and surface. Among deep learning-based methods, V2C-Flow has again the lowest error (0.43 \interval{0.35}mm on average over all landmarks), lagging just marginally behind FreeSurfer v7.2 (FS72) (0.42 \interval{0.39}mm). These results confirm the robustness of V2C-Flow in regions of abnormal myelination and affirm the benefit of the joint training with the proposed feature propagation between WM and pial surfaces via virtual edges. 
}

\subsection{Generalization to new data}
\begin{table}[htb]
\small
\setlength{\tabcolsep}{6.5pt} 
\caption{
Evaluation on external test data. We report the average symmetric surface distance (ASSD) and the 90-percentile Hausdorff distance (HD$_{90}$) and standard deviations over the test set with respect to FreeSurfer surfaces. Values are in mm and averaged over all four cortical surfaces. Numbers in brackets indicate the number of samples in the respective dataset. $^*$For Mindboggle training and test sets, we report the average of a 5-fold cross-validation with 20 test scans per fold to account for the small dataset size ($n=100$).
}

\begin{tabular}{llcc}
    \toprule

    Training set & Test set & ASSD & HD$_{90}$\\
    \midrule
    
    ADNI ($1154$) & J-ADNI ($101$) & 0.236 \interval{0.071} & 0.496 \interval{0.155} \\

    J-ADNI ($350$) & J-ADNI ($101$) & 0.269 \interval{0.092} & 0.580 \interval{0.190}\\
     \addlinespace[.3em]

    ADNI ($1154$) & Mindb. ($100$) & 0.230 \interval{0.052}  & 0.494 \interval{0.123} \\

    Mindb. ($75$) & Mindb. ($20$)$^*$ & 0.258 \interval{0.045} & 0.596 \interval{0.120} \\

    \addlinespace[.3em]
    
    ADNI ($1154$) & OASIS ($80$) & 0.255 \interval{0.075} & 0.554 \interval{0.190}\\

    OASIS ($292$) & OASIS ($80$) & 0.215 \interval{0.038} & 0.485 \interval{0.097} \\

    \bottomrule
\end{tabular}
\label{tab:robustness}
\end{table}

Previous deep learning-based methods for cortex reconstruction have mainly been evaluated on a single dataset. 
However, the generalization to unseen data is crucial for deploying deep learning models to clinical practice. 
In this section, we evaluate the generalization of V2C-Flow to external datasets using the smoothed fsaverage template. To this end, we apply V2C-Flow models trained on ADNI to MRI scans from the J-ADNI~\citep{Iwatsubo2010_JADNI}, the OASIS~\citep{oasis}, and the Mindboggle~\citep{kleinMindboggle2012} studies. 
Supplementary Table~1 reports demographics and acquisition information of the different datasets.
For comparison, we train separate V2C-Flow models on each of these datasets.

\Cref{tab:robustness} reports the accuracy of surface reconstruction on J-ADNI, Mindboggle, and OASIS, when V2C-Flow was either trained on the respective datasets or ADNI. 
These results affirm  that V2C-Flow generalizes well to new data when trained on a large and diverse neuroimaging study like ADNI. If the domain shift is small, e.g., for J-ADNI, which has a nearly identical acquisition protocol to ADNI and a similar age distribution, or the training set is small, e.g., Mindboggle, the model transferred from ADNI even outperforms specialized models trained on the respective dataset in terms of surface accuracy. OASIS, on the other hand, is a homogenous study (only one scanner type) with a considerably lower mean age compared to ADNI (52.7 vs.~72.6 years). 
In this case, training V2C-Flow specifically on OASIS slightly improves the reconstruction accuracy.

\begin{figure*}
    \centering
    \includegraphics[width=\textwidth]{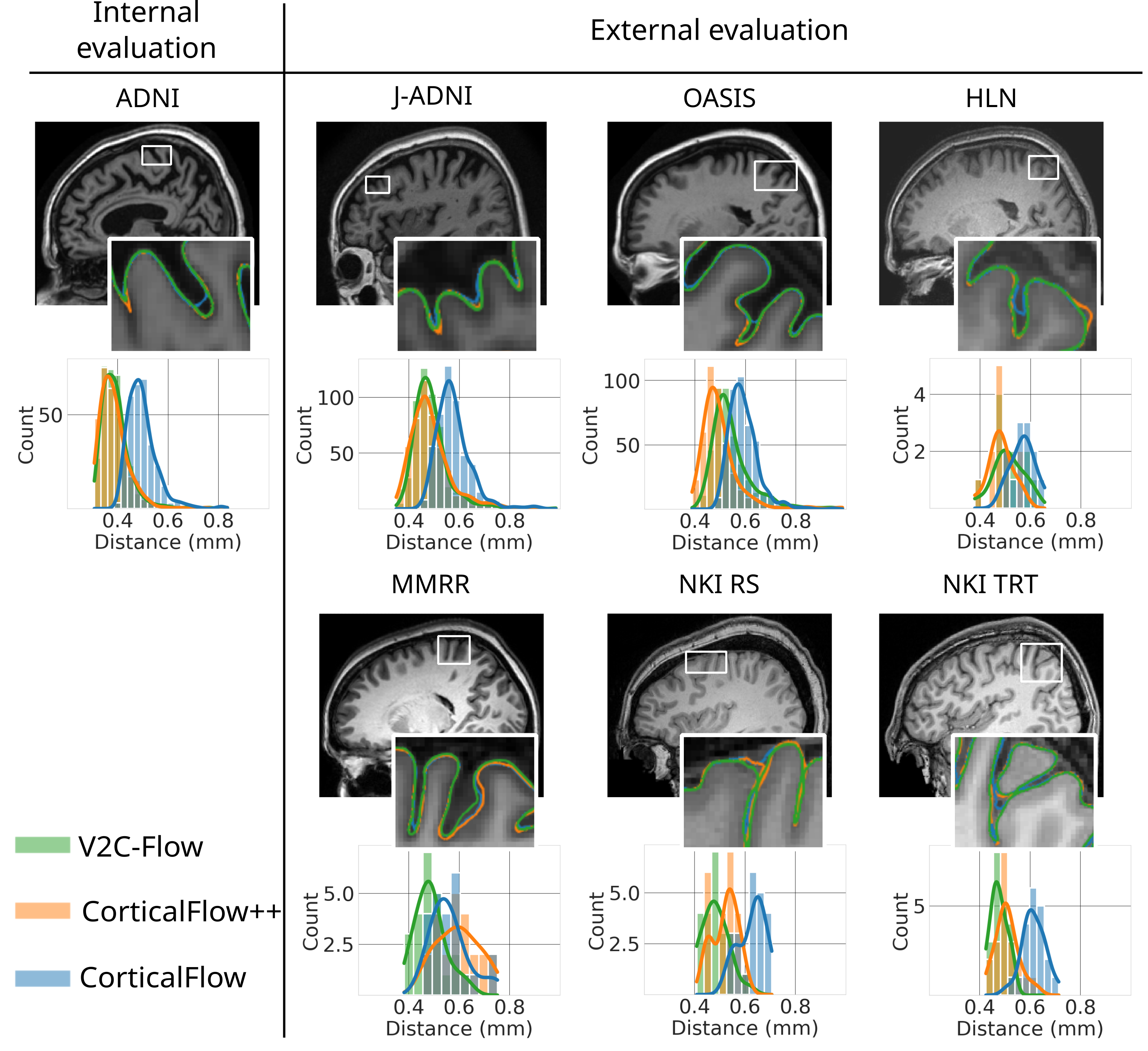}
    \caption{\rev V2C-Flow generalizes well to external datasets. We compare V2C-Flow and CorticalFlow(++) models trained on ADNI using the held-out ADNI test set~($n=323$), J-ADNI~($n=502$), OASIS~($n=416$), HLN~($n=12$), MMRR~($n=23$), NKI-RS~($n=22$), and NKI-TRT~($n=20$) studies. We show exemplary sagittal slices of MRI scans from the respective study and pial surfaces extracted by either method. Below the slices, we plot the histogram of observed HD$_{90}$ distances and corresponding kernel density estimates over the entire datasets.}
    \label{fig:robustness}
\end{figure*}

In \Cref{fig:robustness}, we compare the generalization ability of V2C-Flow to results obtained in a similar setup with CorticalFlow (CF)~\citep{lebrat2021corticalflow}{\rev, and CorticalFlow$^{++}$ (CF$^{++}$)~\citep{santacruz2022_corticalflow++}. CF and V2C-Flow are implemented to be aptly comparable, using the same template, UNet architecture, and optimizer (cf.~implementation details in Supplementary Material). CF$^{++}$ uses a larger template and higher-order ODE solver, which is currently not applicable to V2C-Flow due to training memory and time limitations, but we include it nonetheless as the state-of-the-art baseline.} Quantitatively, we illustrate the distribution of HD$_{90}$ distances for the {\rev three} methods in different evaluation scenarios. In addition, we depict exemplary MRI slices with pial boundaries from all methods. 
Again, we use data from J-ADNI, OASIS, and Mindboggle. However, this time we consider the individual substudies from Mindboggle separately, namely HLN, MMRR, NKI-RS, and NKI-TRT. 
As a reference, we plot the distribution of surface distances on the ADNI test set, which corresponds to an internal evaluation. 
The remaining plots show the performance using the external datasets. 
{\rev We find that V2C-Flow yields higher accuracy than CF on all external test sets and outperforms CF$^{++}$ on MMRR and NKI data. Remarkably, these are also the scans that differ most significantly from the ADNI scans in terms of brightness and contrast. The zoomed-in regions in \Cref{fig:robustness} indicate high accuracy of V2C-Flow in tightly folded regions with fewer vertices than CF$^{++}$. This is rendered possible by the special emphasis on these regions in our curvature-weighted Chamfer loss.}

\subsection{Consistency of reconstructed points} \label{sec:correspondences}
\begin{figure*}
    \centering
    \includegraphics[width=\textwidth]{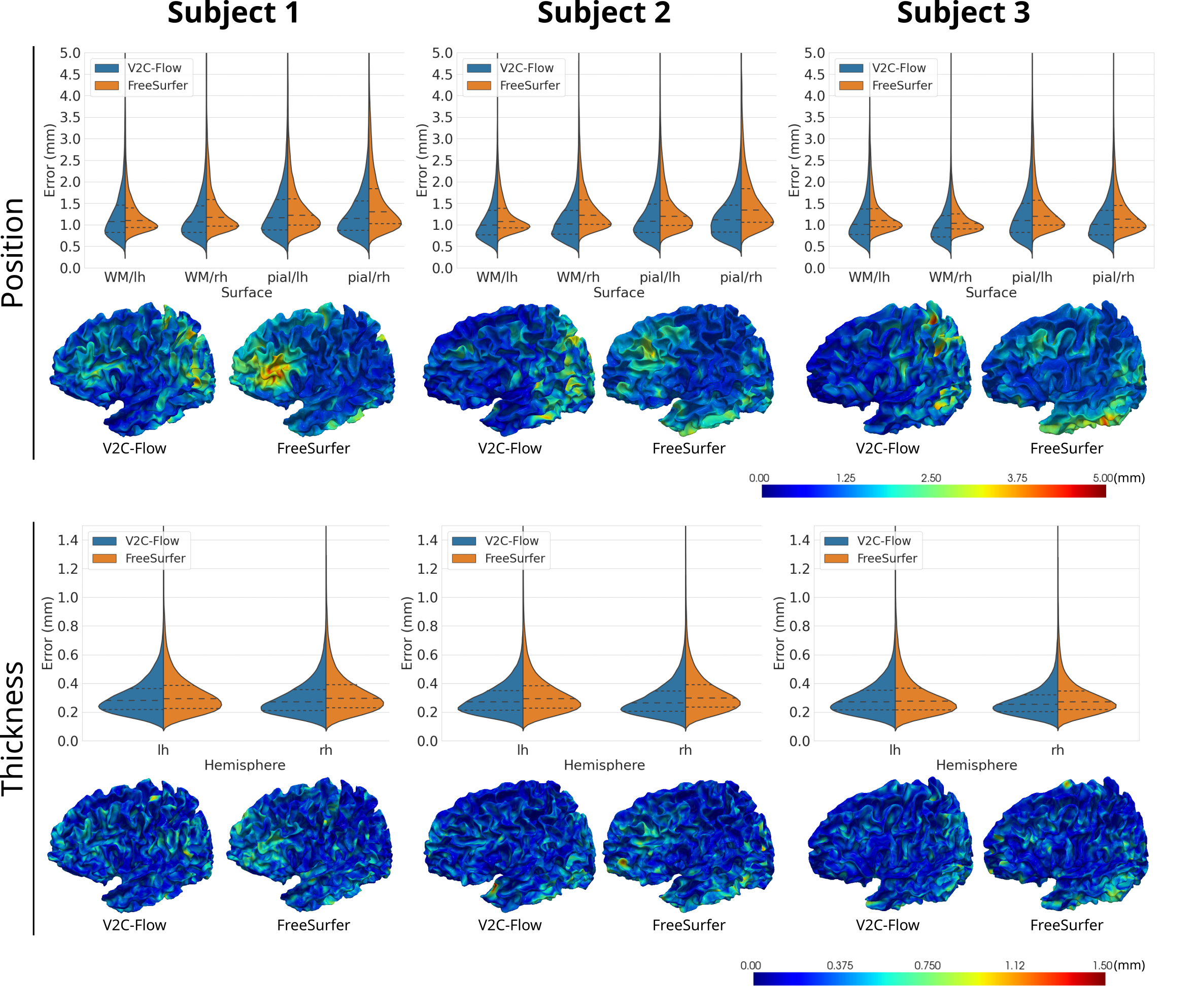}
    \caption{\rev Analysis of the consistency of correspondences with respect to the position (top) and thickness (bottom) on the test-retest dataset. We plot the root-mean-square deviation (RMSD) of vertex positions per surface and thickness measurements per hemisphere from the respective average in the 40 scans per subject (ignoring thickness measures from the "undefined" middle region).
    Violin plots represent kernel density estimates of the per-vertex values for V2C-Flow and FreeSurfer {\revv (v7.2)} with horizontal lines at respective quartiles. Surface renderings show the RMSD per vertex for left WM surfaces. All values are in mm. }
    \label{fig:trt}
\end{figure*}

The results in the previous sections validated the applicability of V2C-Flow for cortical surface reconstruction. Now, we evaluate the consistency of reconstructed points. It is an essential prerequisite for the surfaces to be comparable on a per-vertex basis, ultimately enabling the study of anatomical changes over time or among different populations. Intuitively, the same template vertex should be moved to the same location when provided with different scans of the same subject (given that the scans were acquired within a short period of time to be not susceptible to structural changes). Although this potential represents a major advantage of mesh-based surface reconstruction methods over segmentation- or SDF-based approaches~\citep{groueix2018_3dcoded}, it has not been explored nor validated for cortex reconstruction so far. We close this gap and evaluate the geometric correspondence of reconstructed points and accompanied cortical thickness measurements in V2C-Flow using a public test-retest (TRT) dataset~\citep{maclaren2014} that contains 40 scans from three subjects, respectively, acquired within one month. {\rev We compare the results to FreeSurfer~\citep{fischl1999cortical}, which implements the standard approach for vertex-wise comparison of different cortical meshes}. In a nutshell, FreeSurfer inflates the surfaces to a sphere, registers them based on curvature, and resamples the surfaces to have the same number of vertices. We use the standard fsaverage template for both methods.

On the left WM surfaces in \Cref{fig:trt}, we plot the root-mean-square deviation (RMSD) of vertex positions and thickness measurements per vertex from the respective mean values for each subject in the TRT dataset. In addition, we show 
{\rev violin plots and quartiles of the RMSD per hemisphere for all surfaces}. It can be observed that the consistency of extracted points in terms of RMSD is slightly better in V2C-Flow
compared to the registration-based FreeSurfer pipeline. This is in particular remarkable since we used the raw V2C-Flow output without registration or resampling. Similarly, the cortical thickness measurements associated with each vertex are slightly more consistent in V2C-Flow than in FreeSurfer in terms of RMSD from the average thickness per vertex. 
It stands out that the variation in vertex locations is considerably higher (up to 5mm) than the variation in thickness measurements (up to 1.5mm). 
A likely source for this difference is variations in the scanning procedure that were not compensated by registering the MRI scans to MNI space.

\subsection{Surface parcellation}
\label{sec:parcellation}

The parcellation of the cortex according to a reference atlas, e.g., the popular DKT atlas~\citep{desikan2006,kleinMindboggle2012}, supports region-wise analyses. More precisely, it enables a fine-grained evaluation of morphological measurements like cortical thickness in individual brain regions and draws connections between structural changes and brain functionality. 
In this work, we propose and evaluate two different ways of creating a parcellation of V2C-Flow surfaces (see Supplementary Figure~4 for a visualization of both methods). First, we exploit the learned correspondences between the folded fsaverage template and predicted surfaces by mapping vertex classes directly from the template atlas onto the predicted mesh. This can be done immediately, i.e., without any further processing, and it is therefore the fastest option. We call this method \emph{direct mapping}. Second, we exploit the correspondences between the predicted surfaces to the spherical representation of the fsaverage template. These correspondences allow for curvature-based registration of the predicted surface to the template sphere as implemented in FreeSurfer's \emph{mris\_register} without having to inflate each predicted surface separately (which makes up a large amount of the processing time in FreeSurfer). We call this method \emph{registration-based}.
We assess the accuracy of the proposed methods on the ADNI test set with respect to the FreeSurfer DKT parcellation in terms of Dice overlap. As a baseline, we use FastSurfer {\rev v1.1.2}~\citep{henschel2020} in this case. 

For the mapping-based approach, we obtain an average Dice coefficient (weighted according to region sizes) of $0.82 \ (\text{SD}=0.030)$ and $0.83 \ (\text{SD}=0.029)$ for left and right hemispheres, respectively. This is surprising given that we did not optimize for this task at training time nor did we use any post-processing. However, if a higher parcellation accuracy is required, the quality can be further improved to an average Dice of $0.89 \ (\text{SD}=0.020)$ and $0.89 \ (\text{SD}=0.023)$ per hemisphere by using the registration-based parcellation approach. 
This approach outperforms FastSurfer, which yields Dice scores of $0.87 \ (\text{SD}=0.015)$ and $0.88 \ (\text{SD}=0.022)$ for left and right hemispheres, respectively. 
In addition, we observe in Supplementary Figure~5 that the misclassification of vertices occurs exclusively at the parcel boundaries, which are typically most susceptible to inter-rater variability. That is, the generation of anatomically implausible ``islands'' is effectively prevented.
We conclude that it is straightforward to perform brain parcellation with V2C-Flow, since it maps a template to individual surfaces and not vice-versa, which is in contrast to most existing methods.
Both of the proposed parcellation methods eliminate the need for intricate surface inflation by exploiting learned correspondences to the input template. In practice, the question of which method to choose boils down to an accuracy-speed trade-off: the mapping-based method is as fast as it can get while the registration-based approach yields higher accuracy --- surpassing FastSurfer in terms of parcellation Dice.

\subsection{Analysis of cortical thickness}
\begin{figure*}[!ht]
    \centering
    \begin{subfigure}[b]{0.98\textwidth}
        \includegraphics[width=\textwidth]{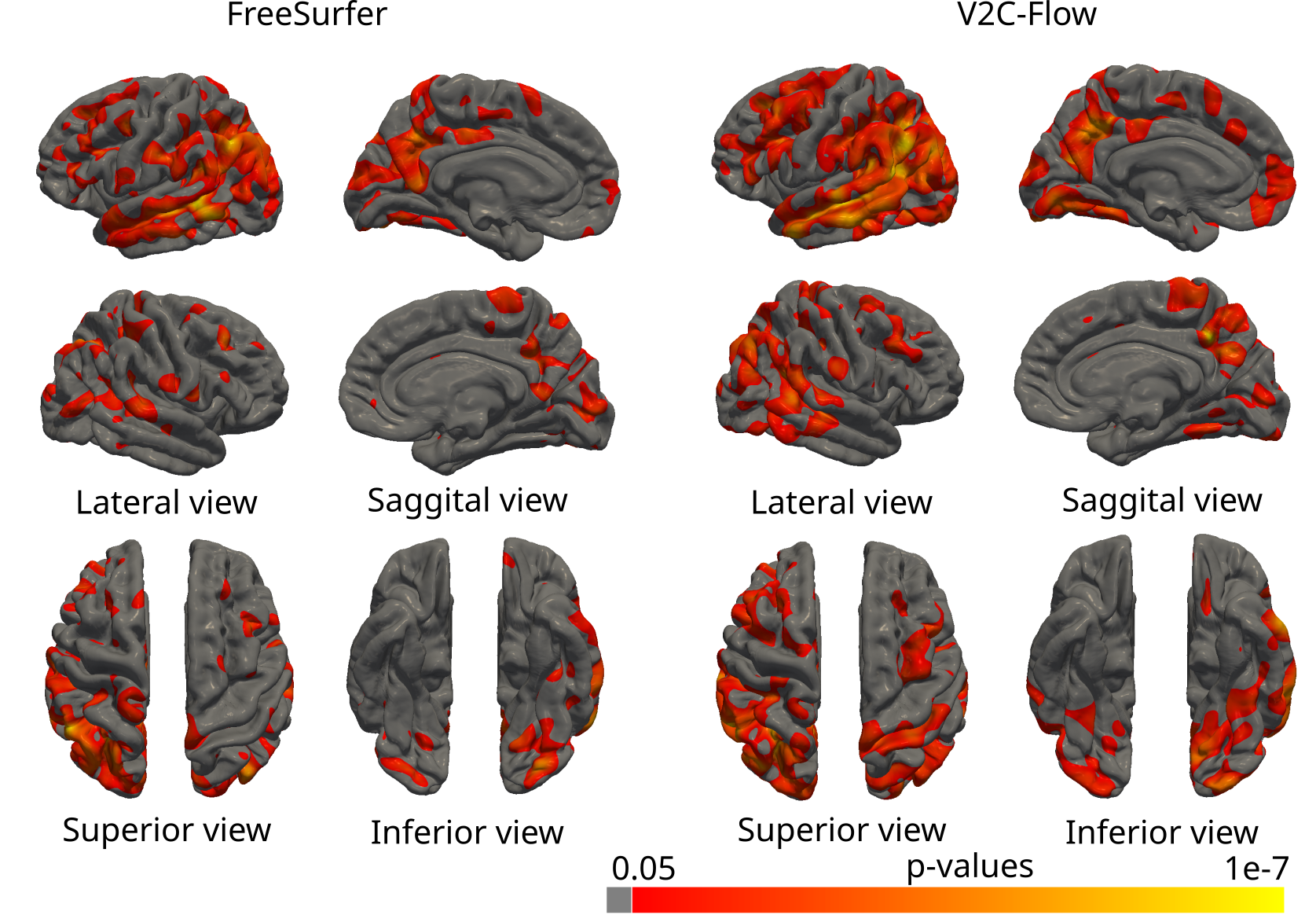}    
        \caption{}
        \label{fig:group_study}
    \end{subfigure}
      \begin{subfigure}[b]{0.6\textwidth}
        \includegraphics[width=\textwidth]{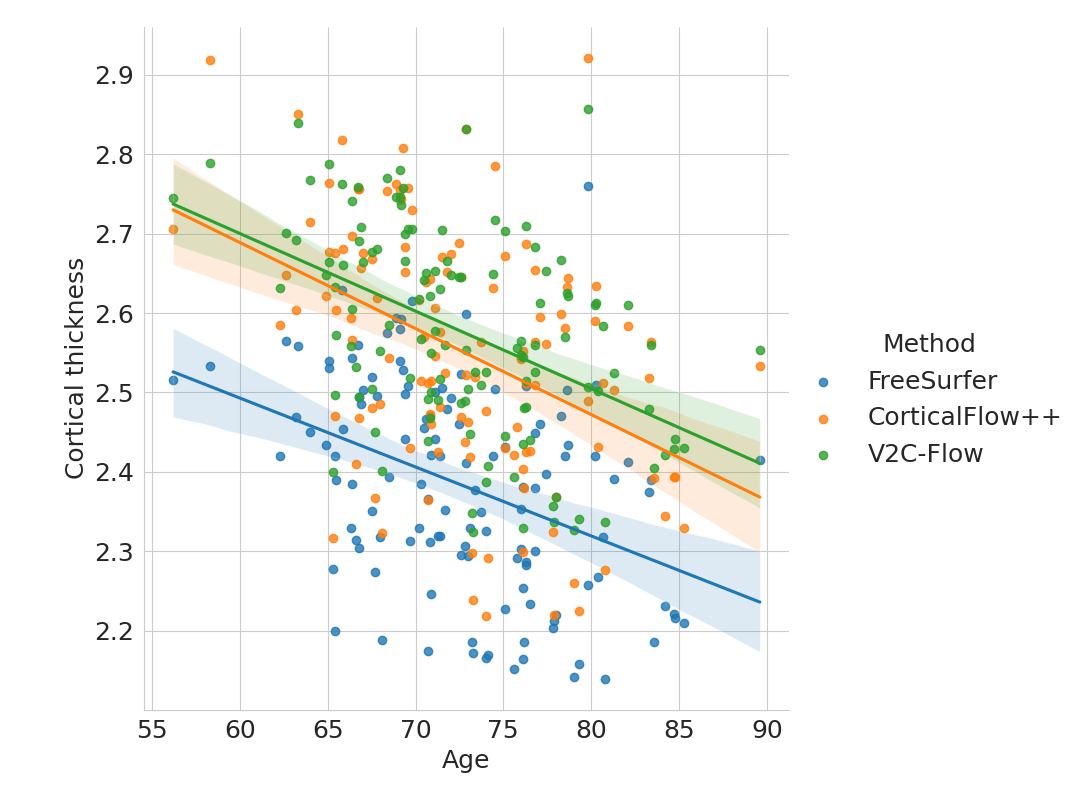}    
        \caption{}
        \label{fig:thickness-age}
    \end{subfigure}
       \begin{subfigure}[b]{0.38\textwidth}
        \includegraphics[width=\textwidth]{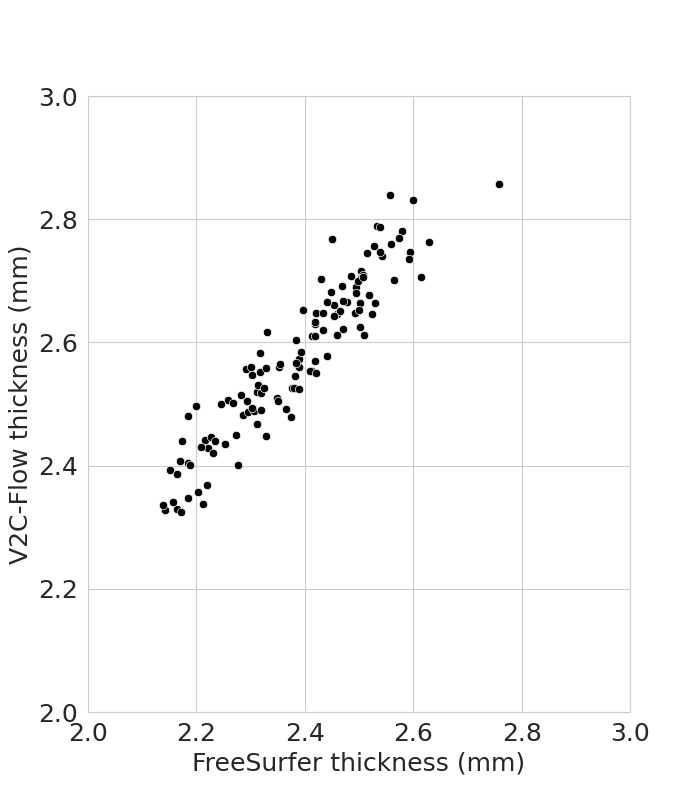}    
        \caption{}
        \label{fig:thickness-corr}
    \end{subfigure}

    \caption{\rev Analysis of cortical thickness measurements from V2C-Flow. 
    (a) Group study of per-vertex cortical thickness in patients diagnosed with AD (male: $n=28$, female: $n=22$) and healthy controls (male: $n=53$, female: $n=71$). Colors represent uncorrected p-values that indicate a significantly lower cortical thickness in AD subjects (one-sided t-test, general linear model with respect to age and sex) for FreeSurfer (left) and V2C-Flow (right). (b) Scatter plot of cortical thickness with respect to age for FreeSurfer {\revv (v7.2)}, V2C-Flow, and CorticalFlow$^{++}$. Lines show ordinary linear regression fits and 95\% confidence intervals. (c) Correlation of mean cortical thickness per subject obtained from FreeSurfer and V2C-Flow ($r=.94$). Plots (b) and (c) are based on the right hemispheres of healthy subjects from the ADNI test set ($n=124$).}
    \label{fig:thickness}
\end{figure*}

A key application for extracted cortical surfaces is the computation of cortical thickness. 
This allows for a detailed study of brain atrophy due to aging or neurodegenerative diseases like Alzheimer's disease (AD). 
Using the ADNI test set, we extract cortical thickness measurements from V2C-Flow and analyze them in relation to diagnosis and age. We compute the cortical thickness per WM vertex from the nearest point on the pial surface.

\Cref{fig:group_study} highlights regions with significantly (t-test, one-sided) lower cortical thickness in patients diagnosed with AD (male: $n=28$, female: $n=22$) compared to the healthy control group (male: $n=53$, female: $n=71$) on the ADNI test set. To this end, we fitted a general linear model to the vertex-wise thickness measurements on the fsaverage template --- taking into consideration the age and sex of the subjects. 
The plots in \Cref{fig:group_study} show that V2C-Flow and FreeSurfer indicate similar cortical regions that are affected by AD-related atrophy, with slightly higher significance in tests for V2C-Flow. 
Importantly, thickness measures in V2C-Flow can be compared directly, while FreeSurfer surfaces need to be inflated, registered, and re-sampled beforehand. This reduces the computation time of the sole group comparison (excluding surface extraction) from several minutes to a few seconds.

{\rev
Besides the observed atrophy due to Alzheimer's disease, we also investigate the age-related trajectory of mean cortical thickness in healthy individuals. To this end, we compare measures obtained from FreeSurfer, CorticalFlow$^{++}$~(CF$^{++}$), and V2C-Flow surfaces in \Cref{fig:thickness-age}. For best comparison, we used the same thickness estimation in all methods. We excluded thickness measures from the "undefined" region in the DKT atlas, i.e., the region that separates left and right cortical hemispheres, since they are commonly not taken into consideration. Since CF$^{++}$ surfaces do not come with a parcellation, we inflated, registered, and parcellated them with FreeSurfer tools.
The slopes for FreeSurfer, V2C-Flow, and CF$^{++}$ are \mbox{-8.7{\textpm}1.8{\textmu}m}, -9.8{\textpm}1.7{\textmu}m, and -10.8{\textpm}2.1{\textmu}m annual change of cortical thickness, respectively. 
The slope of V2C-Flow is within the standard error of the FreeSurfer estimate, which is not the case for CF$^{++}$. 
The intercepts are 3.01{\textpm}0.13mm, 3.29{\textpm}0.12mm, and 3.34{\textpm}0.15mm for FreeSurfer, V2C-Flow, and CF$^{++}$, respectively. 
The offsets for the deep learning-based methods (CF$^{++}$ and V2C-Flow) are slightly higher than for FreeSurfer. While this difference calls for further investigation, the relative change in thickness is of main interest for the majority of applications, which relates to the slope.  
In general, we observed a strong correlation ($r=.94$) between the thickness measurements of V2C-Flow and FreeSurfer, cf.~\Cref{fig:thickness-corr}. In comparison, the correlation between CorticalFlow$^{++}$ and FreeSurfer is $r=.93$. This analysis is based on the average cortical thickness of the right hemisphere considering all healthy subjects from the ADNI test set ($n=124$).

}

\section{Limitations}

Different to most other applications of deep learning, the creation of a large gold standard in the sense of human-annotated target meshes is not practicable for cortical surface extraction from MRI. Hence, for training our models as well as for their evaluation, we rely on FreeSurfer~\citep{fischl2012freesurfer} (version 7.2). This has become a standard practice~\citep{santacruz2021,lebrat2021corticalflow,Bongratz_2022_CVPR} since FreeSurfer has been validated extensively in a multitude of settings~\citep{schwarz_large-scale_2016,cardinale_validation_2014,popescu_postmortem_2016}. Nonetheless, FreeSurfer can extract erroneous surfaces for some scans; see \Cref{fig:comparison} and Supplementary Figure~6. Although we tried to remove such scans, there might still be noisy labels in our datasets. 
It further needs to be considered that surfaces from V2C-Flow can contain self-intersections, which may need to be removed by post-processing if a specific application has an according requirement.  
In our experiments on several datasets, we have included scans from healthy subjects as well as subjects with cognitive impairment,  Alzheimer's disease{\rev, and multiple sclerosis}. 
The surface reconstruction in the presence of other brain pathologies, e.g., tumors, has yet to be assessed. 
Still, we believe that the learning-based V2C-Flow can extend to such cases with the appropriate composition of the training set and that the current model already allows for a wide range of applications. Since our models are trained with FreeSurfer's fsaverage template, V2C-Flow and FreeSurfer surfaces can be compared in practice. That is, point- or region-wise measurements such as cortical thickness can be compared based on fsaverage as a common basis, no matter which of the two methods was used for the surface extraction. 
This also allows for seamless integration of the V2C-Flow output in the rich toolbox of FreeSurfer for cortex analyses.

\section{Conclusion}
We have introduced Vox2Cortex-Flow (V2C-Flow), a novel end-to-end deep learning method for cortical surface reconstruction that maintains vertex-wise correspondences to an input template. To this end, a continuous neural mesh-deformation scheme based on stacked graph NODEs and conditioned on deep image features is proposed. Our extensive experiments on internal and external neuroimaging data show that V2C-Flow reconstructs all four cortical surfaces simultaneously within less than two seconds on a standard GPU, yields state-of-the-art surface accuracy, and avoids topological errors in related methods. Moreover, we find that the consistency of reconstructed points, learned by our curvature-weighted Chamfer loss, and accompanied cortical thickness measurements are surprisingly higher than in the spherical-registration-based baseline. This observation challenges the common approach of segmentation-based methods that separate surface extraction from the registration. Instead, we suggest a consolidated approach that combines segmentation, topology correction, surface extraction, inflation, registration, and resampling into a single forward pass of our geometric neural network.

\section*{Declaration of competing interest}
The authors declare that they have no known competing financial interests or personal relationships that could have appeared to
influence the work reported in this paper.

\section*{Data availability}
The code is made publicly available on github with the link mentioned in the manuscript and used datasets are public.

\section*{Acknowledgments}
This research was supported by the Bavarian State Ministry of Science and the Arts and coordinated by the Bavarian Research Institute for Digital Transformation, and the Federal Ministry of Education and Research in the call for Computational Life Sciences (DeepMentia, 031L0200A), and the German research foundation. 
The authors gratefully acknowledge the Leibniz Supercomputing Centre (\url{www.lrz.de}) for providing computational resources. Thanks also to Carl Winkler for running CortexODE and to Kathleen Larson for helping with the JHU landmarks.

\bibliographystyle{model2-names.bst}\biboptions{authoryear}
\bibliography{refs}

\clearpage

\renewcommand{\figurename}{Supplementary Fig.}
\renewcommand{\tablename}{Supplementary Table}
\setcounter{figure}{0}  
\setcounter{table}{0}  
\setcounter{equation}{0}  

\section*{Supplementary Material}

\subsection*{Surface distances}
For the evaluation of surface accuracy, we use the average symmetric surface distance (ASSD) and a symmetric 90-percentile Hausdorff distance (HD$_{90}$). Given two surfaces $\mathcal{M}_1, \mathcal{M}_2 \subset \mathbb{R}^3$, we sample $N=100,000$ random points from each surface, i.e., two sets $\mathcal{Q}_1=\{q_i \mid q_i \in \mathcal{M}_1 \wedge 1\leq i \leq N\}$, $\mathcal{Q}_2=\{q_i \mid q_i \in \mathcal{M}_2 \wedge 1\leq i \leq N\}$. Then, we calculate the ASSD as
\begin{equation}
\begin{split}
    d_{\text{ASSD}} (\mathcal{M}_1, \mathcal{M}_2) 
    &= \frac{1}{\abs{\mathcal{Q}_1} + \abs{\mathcal{Q}_2}} \left[ \sum_{q_i \in \mathcal{Q}_1} \underset{q_j \in \mathcal{M}_2}{\min} d(q_i, q_j) \right. \\
    &+ \left. \sum_{{\rev q}_i \in \mathcal{Q}_2} \underset{q_j \in \mathcal{M}_1}{\min} d(q_i, q_j) \right] , 
\end{split}
\end{equation}
where $d(x, y)$ is the Euclidean distance between two points $x,y \in \mathbb{R}^3$. The HD$_{90}$ calculates as
\begin{equation}
\begin{split} 
    \text{HD}_{90}(\mathcal{M}_1, \mathcal{M}_2) = \max \{
    & P_{90}(\{\underset{q_j \in \mathcal{M}_2}{\min} d(q_i, q_j) \mid q_i \in \mathcal{Q}_1\}), \\
    & P_{90}(\{\underset{q_j \in \mathcal{M}_1}{\min} d(q_i, q_j) \mid q_i \in \mathcal{Q}_2\})
    \},
\end{split}
\end{equation}
with $P_{90}(\cdot)$ denoting the 90-percentile of the respective values.

\subsection*{Comparison of surface parcellations} \label{sec:parellation-computation}
In order to compare predicted parcellations, i.e., vertex classifications, to (pseudo-)ground-truth parcellations on non-identical two-dimensional manifolds, we used a bi-directional nearest-neighbor approach. That is, we computed the average of the parcellation Dice on either surface by comparing the predicted (respectively ground-truth) class of a point to the ground-truth (respectively predicted) class of the nearest point on the other surface. Here, we used the central surfaces lying in the middle of WM and pial boundaries to avoid a corruption of the result from using either the WM or pial surface. This is meaningful since the parcellation of the WM and the pial surfaces should be in analogy to each other, i.e., the assigned class of a WM surface matches the class of the corresponding vertex on the pial surface.

\subsection*{Statistics for cortical thickness-based comparison of healthy controls with AD patients} \label{sec:group-statistics}
We calculated identical general linear regression models to detect group differences between patients diagnosed with Alzheimer's disease (AD) and a healthy control (HC) group (dependent variable). To this end, we used FreeSurfer's \emph{mri\_glmfit} and  defined the contrast such that a difference between AD and HC subjects is detected, while accounting for age- and sex-related variations in the data. Note that the resulting p-values from V2C-Flow and FreeSurfer are comparable since they were calculated from the same input data, i.e., the same sample sizes~\citep{henschel2020}. As input to the regression model, we employed vertex-wise thickness measures on the fsaverage template, smoothed with a full-width-half-max kernel of 10mm. For FreeSurfer, we mapped thickness measures from individual surfaces to the template via \emph{mris\_preproc} prior to the analysis. In the case of V2C-Flow, we used the learned correspondences between the template and the individual surfaces, hence omitting the resampling. 

\subsection*{Implementation of baseline methods}
Our implementation of DeepCSR is based on the original repository\footnote{\url{https://bitbucket.csiro.au/projects/CRCPMAX/repos/deepcsr/browse}}. We sample points in DeepCSR on a grid with spacing of 0.5mm (voxel resolution is 1mm) and closely follow the pre-processing described in~\citep{santacruz2021}.  
For CorticalFlow$^{++}$~\citep{santacruz2022_corticalflow++}{\rev, TopoFit~\citep{hoopes2022topofit}, CortexODE~\citep{ma2022_cortexODE},} and FastSurfer {\rev v1.1.2}~\citep{henschel2020}, we applied the code and template as published in the original repositories\footnote{\url{https://bitbucket.csiro.au/projects/CRCPMAX/repos/corticalflow/browse}}\footnote{\url{https://github.com/ahoopes/topofit}}\footnote{\url{https://github.com/m-qiang/CortexODE}}\footnote{\url{https://github.com/deep-mi/FastSurfer}}. {\rev TopoFit was extended for pial surfaces and each of the four models was trained for 500 epochs (vs.~105 epochs for the joint model in V2C-Flow).} CorticalFlow~\citep{lebrat2021corticalflow} was adapted to be aptly comparable to V2C-Flow {\rev for best evaluation of the different architectures}. In particular, we used the same optimizer, template, Euler integration scheme, and UNet architecture as in V2C-Flow and predicted all four cortical surfaces from a single UNet (with different output channels per deformation field). Similarly, Vox2Cortex~\citep{Bongratz_2022_CVPR} was updated to match the implementation of V2C-Flow for best comparison.

\begin{table*}[t]
\centering
\footnotesize
\caption{Demographic and acquisition information about the datasets used in our experiments.}

\begin{tabular}{lccccr}
     \toprule

     Dataset & Subset & \makecell{Mean age\\ in years } & \makecell{Diagnoses\\($n$ subjects)} & \makecell{Field\\ strength} & Scanner \\
     \midrule
     ADNI && 72.6 (SD=7.0) & \makecell{Healthy ($632$)\\MCI ($756$)\\AD ($259$) }& 1.5T/3T & multi-site\\

        \addlinespace[.3em]
        
     J-ADNI && 72.0 (SD=6.4) & \makecell{Healthy ($145$)\\ MCI ($218$)\\ AD ($139$)} & 1.5T & multi-site\\
   
   \addlinespace[.3em]
   
     OASIS && 52.7 (SD=25.1) & \makecell{Healthy ($316$)\\ mild AD ($70$) \\ moderate AD ($30$)} & 1.5T & Siemens\\

   \addlinespace[.3em]
   
    TRT & & 26, 30, 31 & Healthy ($3\times 40$) & 3T & GE \\

      \addlinespace[.3em]
      
    \multirow{6}{*}{\makecell{Mind-\\boggle}} & HLN & 27.8 (SD=4.62) & Healthy ($12$) & 3T & Philips \\
     & MMRR & 31 (SD=9.2) & Healthy ($23$) & 3T &  Philips \\
     & NKI-RS & 26.0 (SD=5.2) & n/a ($22$) & 3T & Siemens \\
     & NKI-TRT & 31.4 (SD=11.1) & n/a ($20$) & 3T & Siemens \\
     & OASIS-TRT & 23.4 (SD=3.9) & Healthy ($20$) & 1.5T & Siemens \\
     & Other & 41, 41, 36 & n/a ($3$) & n/a & n/a  \\

     \multirow{2}{*}{\rev JHU} & & \rev 39.4 (30-49) & \rev Healthy (5) & \multirow{2}{*}{\rev n/a} & \multirow{2}{*}{\rev n/a} \\
     & & \rev 48.4 (40–59) & \rev MS (5) && \\

    \bottomrule
\end{tabular}
\label{tab:datasets}
\end{table*}

\begin{table*}[h]
\renewcommand\bfdefault{b}
\setlength{\tabcolsep}{3pt} 

    \centering
    \footnotesize
    \caption{\rev Relative number of self-intersecting faces (SIF) within surface meshes and absolute (relative) number of intersecting faces (IF) between WM and pial surfaces. We did not count intersections in the medial wall where WM and pial surfaces usually coincide. We further report average symmetric surface distance (ASSD) in mm. All values are computed over both hemispheres. To fix self-intersections, we used the publicly available implementation of MeshFix~\cite{attene2010}. $^*$CortexODE does not come with correspondences between surfaces. $^{**}$We resampled the FreeSurfer surfaces to the fsaverage template using the \texttt{mri\_surf2surf} command.}
    {\rev
    \begin{tabular}{lccccl}
    \toprule
          & \multicolumn{2}{c}{WM} & \multicolumn{2}{c}{Pial} & \multicolumn{1}{c}{WM $||$ Pial} \\
          \cmidrule(lr){2-3} \cmidrule(lr){4-5} \cmidrule(lr){6-6}
         Method & ASSD & SIF & ASSD & SIF & \multicolumn{1}{c}{IF} \\
         \midrule

         V2C-Flow  &  .178 \interval{.036} & 0.66 \interval{0.23}\% & .175 \interval{.026} & 1.9 \interval{0.87}\% & \textbf{82 \interval{111} (0.01 \interval{0.02}\%)} \\

         \vspace{3pt}
        + Meshfix & .178 \interval{.036} & 0  &  .181 \interval{.026} & 0  &  144 \interval{131} (0.02 \interval{0.02}\%) \\
        
        CF$^{++}$ & .181 \interval{.035} &  \textless$\expnumber{1}{-4}$ & .169 \interval{.031} & 0.04 \interval{0.05}\% &  264 \interval{209} (0.02 \interval{0.02}\%)  \\

        TopoFit & .210 \interval{.036}  & 0.03 \interval{0.04}\% & .232 \interval{.038} & 0.65 \interval{0.46}\% & 396 \interval{300} (0.07 \interval{0.05}\% \\

        CODE$^*$ & .172 \interval{.039} &  \textless$\expnumber{1}{-4}$ & .187 \interval{.035} & 0.09 \interval{0.05}\% & 95 \interval{98} (0.02 \interval{0.02}\%) \\

        FS72$^{**}$ & - & 34 \interval{4.5}\% & - & 41 \interval{4.9}\% & 94 \interval{1104} (0.02 \interval{0.2}\%)\\

         \bottomrule
    \end{tabular}
    }
    \label{tab:intersections}
\end{table*}

\begin{table*}[t]
    \centering
    \setlength{\tabcolsep}{3pt} 

    \small
    \caption{\rev Evaluation of cortex reconstruction methods with respect to manual landmarks from the JHU dataset. We distinguish healthy (HC) subjects and patients diagnosed with multiple sclerosis (MS). Values indicate the mean distance of the landmarks to the mesh surface (\interval{standard deviation}) in mm.}
    {\rev
    \begin{tabular}{lccccccc}
    \toprule
    
        Diagnosis & Surface & Rater & CF++ & CODE & TopoFit & V2C-Flow (S) & FS72  \\ \midrule
        
        \multirow{5}{*}{HC} & \multirow{2}{*}{pial} & A & 
        .69 \interval{.44} & 
        .85 \interval{.46} & 
        .78 \interval{.51} & 
        .57 \interval{.39} & 
        .50 \interval{.37} \\ 
         
         &  & B & 
         .48 \interval{.37} & 
         .57 \interval{.40} &  
        .53 \interval{.40} & 
        .44 \interval{.36} & 
        .43 \interval{.37} \\ 
        
         & \multirow{2}{*}{WM} & A & 
         .42 \interval{.34} & 
         .41 \interval{.33} &
         .45 \interval{.37}& 
         .44 \interval{.36} & 
         .45 \interval{.36} \\ 
         
         &  & B & 
         .44 \interval{.37} & 
         .45 \interval{.38} &  
         .45 \interval{.38} 	& 
         .41 \interval{.33} & 
         .40 \interval{.31} \\

        \multirow{5}{*}{MS} & \multirow{2}{*}{pial} & A & 
        .56 \interval{.37} & 
        .77 \interval{.46} &  
        .71 \interval{.44}&
        .39 \interval{.27} &
        .37 \interval{.29} \\
        
         &  & B & 
         .42 \interval{.33} & 
         .59 \interval{.46} &  
         .55 \interval{.44}& 
         .33 \interval{.26} & 
         .34 \interval{.34} \\
         
         & \multirow{2}{*}{WM} & A & 
         .41 \interval{.33} & 
         .43 \interval{.42} & 
         .43 \interval{.39} 	& 
         .46 \interval{.37} & 
         .43 \interval{.50} \\ 
         
         &  & B & 
         .42 \interval{.37} & 
         .47 \interval{.48} & 
         .48 \interval{.47} 	& 
         .43 \interval{.35} &
         .42 \interval{.53} \\ 

        \midrule

        \multicolumn{3}{c}{all} &

        .48 \interval{.38} &
        .57 \interval{.45} &
        .55 \interval{.44} &
        .43 \interval{.35} &
        .42 \interval{.39} \\

         \bottomrule
    \end{tabular}
    }
\label{tab:accuracy_no_lesion}

\end{table*}

\begin{table*}[h]
\renewcommand\bfdefault{b}
\setlength{\tabcolsep}{2.pt} 

    \centering
    \footnotesize
    \caption{\revv Ablation study of the proposed training with virtual edges. We trained two models --- with and without virtual edges --- on the ADNI training set with the smoothed fsaverage template. We compare the results using the ADNI validation set and the Mindboggle data by computing the average symmetric surface distance (ASSD) to the FreeSurfer {\revv (v7.2)} meshes and the number of intersecting faces between WM and pial surfaces (WM $||$ Pial IF). Values are the mean (\interval{standard deviation}) over all predictions for both hemispheres.}
    {\revv
    \begin{tabular}{lcccc}
    \toprule
          & \multicolumn{2}{c}{ADNI} & \multicolumn{2}{c}{Mindboggle} \\
          \cmidrule(lr){2-3} \cmidrule(lr){4-5} 
         Method & ASSD (mm) & WM $||$ Pial IF & ASSD (mm) & WM $||$ Pial IF  \\
         \midrule

         V2C-Flow  & .177 \interval{.036} & 90 \interval{159} (.015 \interval{.027}\%) & .230 \interval{.052} & 152 \interval{144} (.025 \interval{.024}\%)   \\
        
        w/o virtual e. & .187 \interval{.036} &  112 \interval{273} (.019 \interval{.046}\%) & .245 \interval{.062}&  152 \interval{131} (.025 \interval{.022}\%)\\
        
         \bottomrule
    \end{tabular}
    }
    \label{tab:my_label}
\end{table*}

\begin{figure*}
    \centering
    \includegraphics[width=\linewidth]{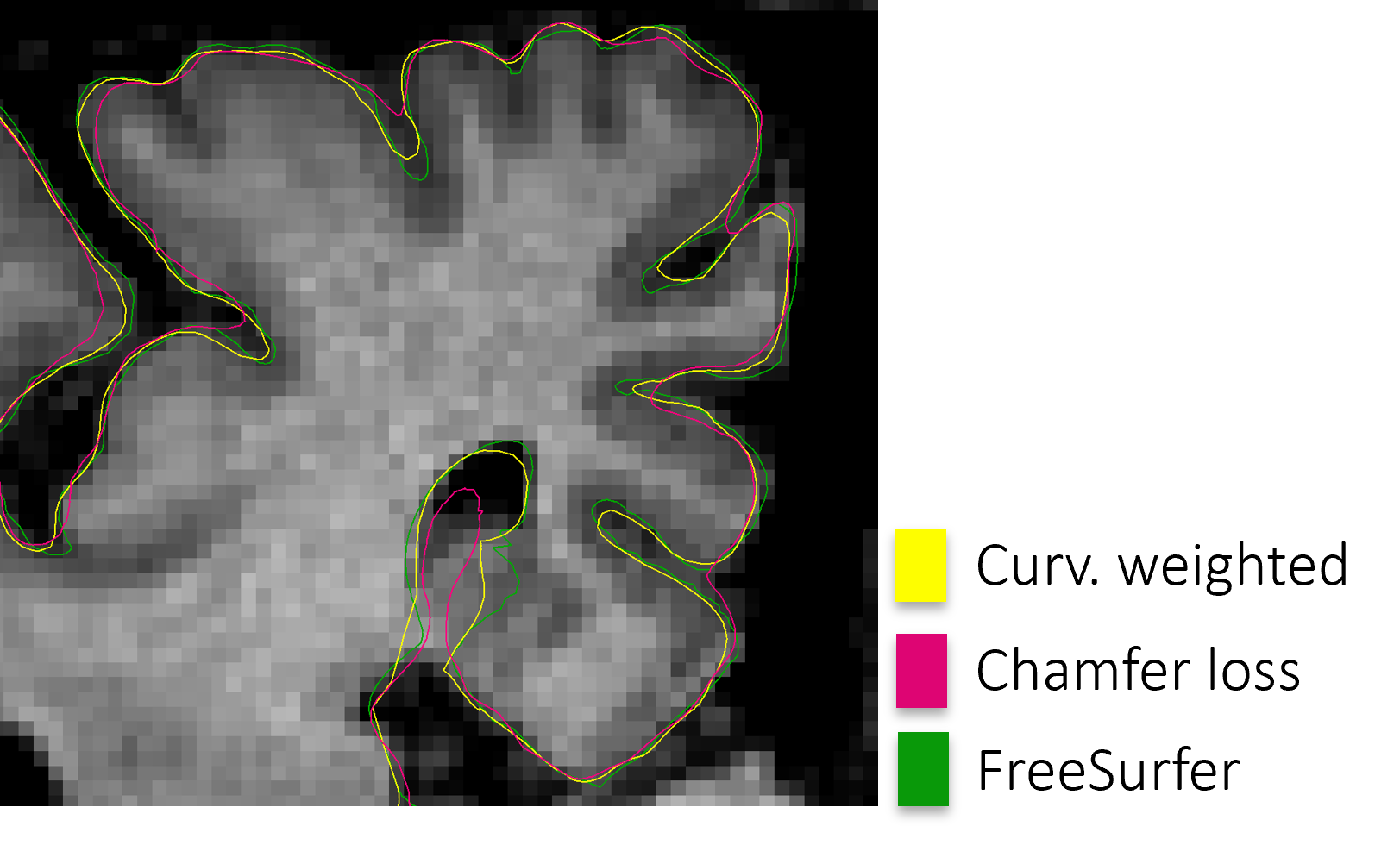}
    \caption{The curvature-weighted Chamfer loss re-weights the ratio between point and regularization loss terms locally based on ground-truth curvature. Thereby, it enhances the accuracy in highly curved regions compared to the standard Chamfer loss (with the same regularization terms). The FreeSurfer surface is shown as a reference.}
    \label{fig:curvature-loss}
\end{figure*}

\begin{figure*}
    \centering
    \includegraphics[width=\textwidth]{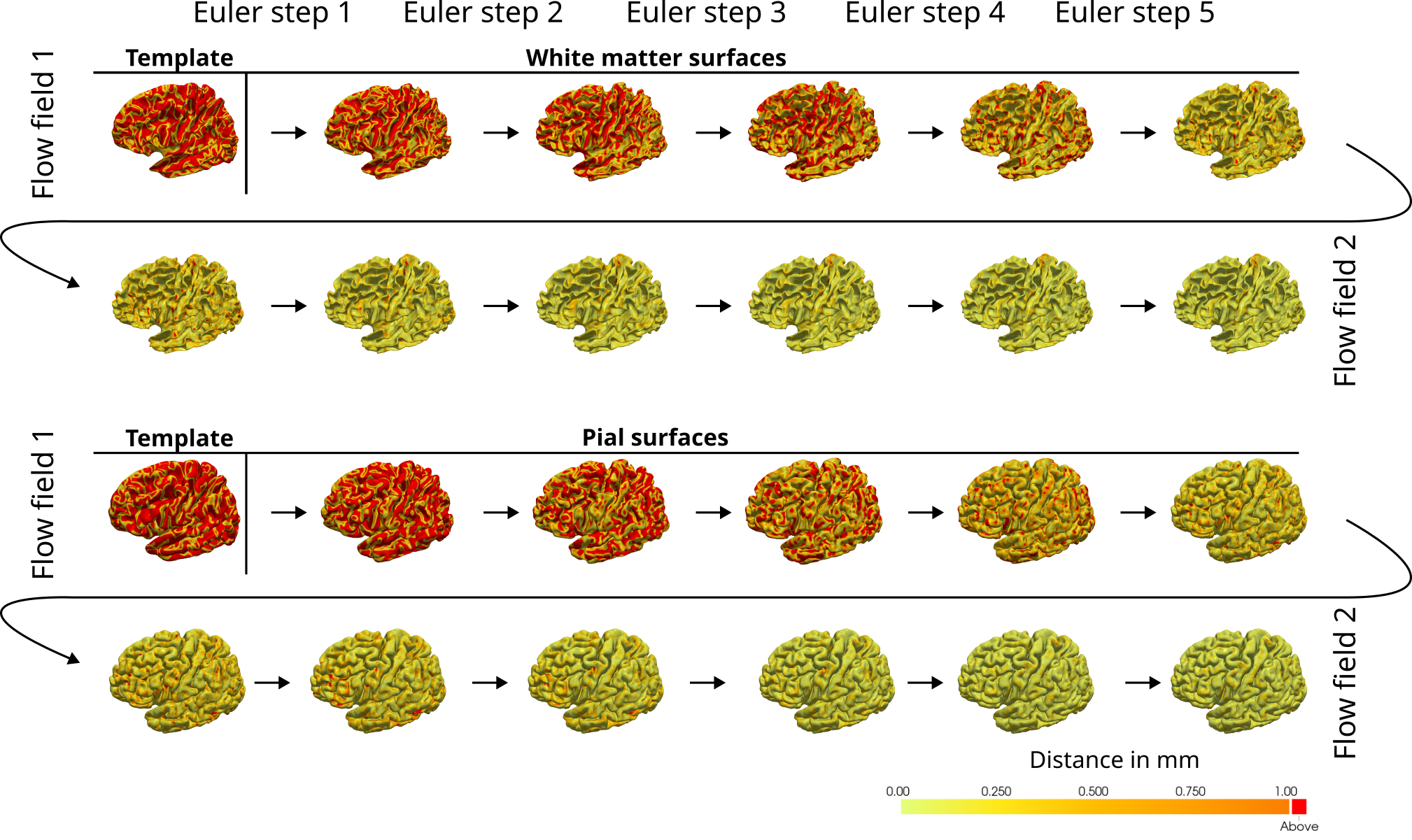}
    \caption{Illustration of the deformation in V2C-Flow from the fsaverage template to the individual white matter (top) and pial (bottom) surfaces with two flow fields ($S=2$). Each flow is integrated numerically with the forward Euler method. We plot the per-vertex distance for white matter and pial surfaces of the left hemisphere to the FreeSurfer silver standard for a sample from the ADNI test set
    .The error decreases with the evolution of the mesh. For an animation of the deformation process see supplemental videos. }
    \label{fig:vis-v2c-flow}
\end{figure*}

\begin{figure*}
 \centering
\begin{subfigure}[b]{0.32\textwidth}
     \includegraphics[width=\textwidth]{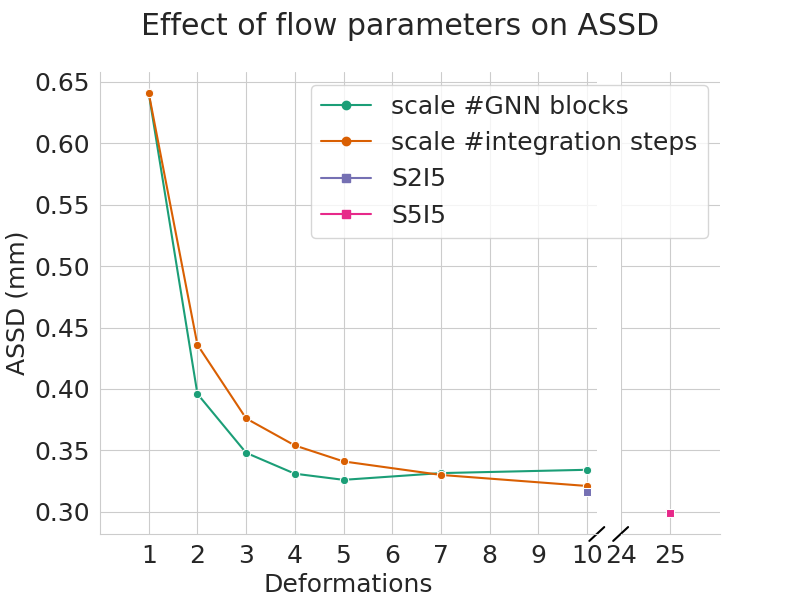}
     \subcaption{}
\end{subfigure}
\begin{subfigure}{0.32\textwidth}
     \includegraphics[width=\textwidth]{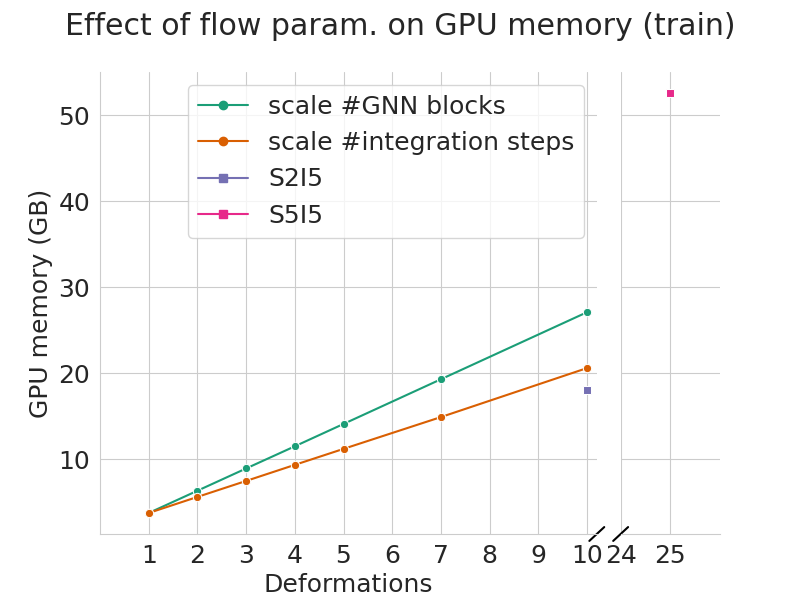}
     \subcaption{}
\end{subfigure}
\begin{subfigure}{0.32\textwidth}
     \includegraphics[width=\textwidth]{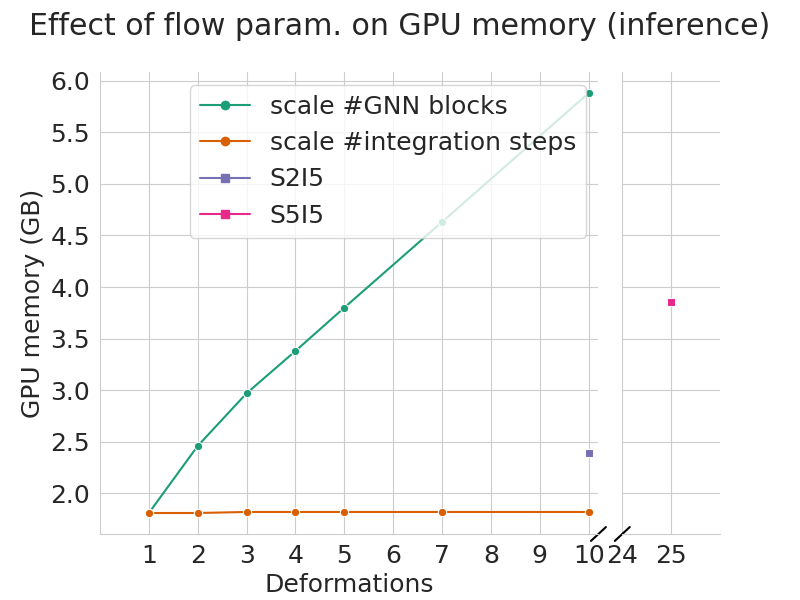}
     \subcaption{}
\end{subfigure}
    \caption{\revv Ablation study of the proposed graph NODEs, i.e., integrable GNN blocks, in V2C-Flow (reduced image and mesh resolution, OASIS data). Compared to stacking non-integrable GNN blocks (green curves), a single graph NODE (orange curves) yields better reconstruction accuracy with $>7$ Euler integration steps at lower memory cost. The best results are obtained with a combination of both, i.e., multiple graph NODEs $S$ and multiple integration steps $I$ per graph NODE, leading to $S \times I$ atomic deformations. Shown are (a) the average symmetric surface distance (ASSD) to the FreeSurfer silver standard,  (b) the required GPU memory at training time, and (c) the required GPU memory at inference time in dependence on the number of atomic deformations.
    }
    \label{fig:gnn_euler_steps}
\end{figure*}

\begin{figure*}
    \centering
\begin{subfigure}[b]{\textwidth}
     \includegraphics[width=\textwidth]{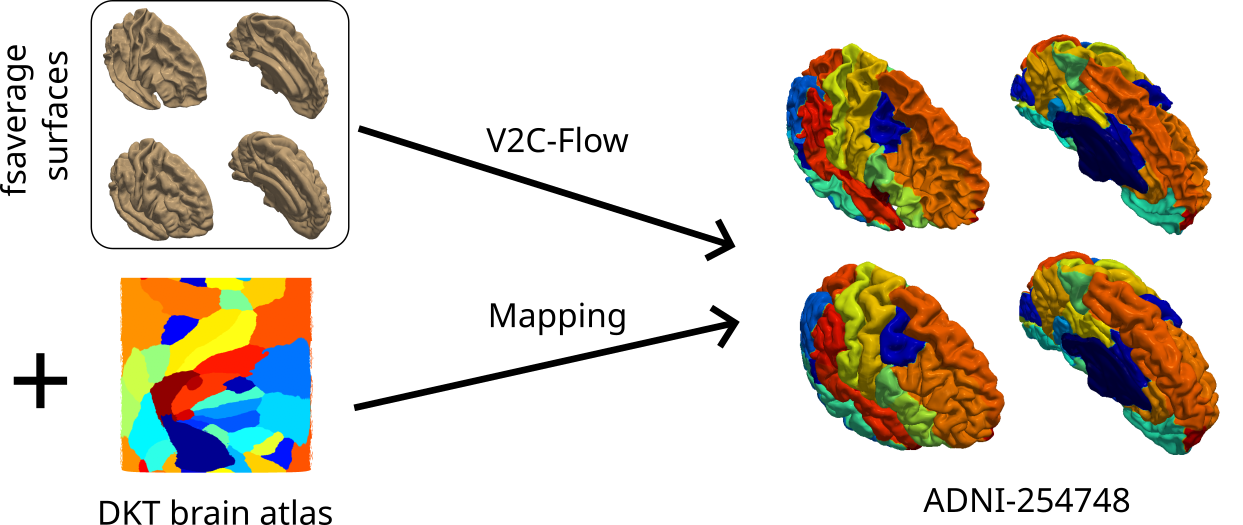}
     \subcaption{}
\end{subfigure}
\begin{subfigure}[b]{\textwidth}
     \includegraphics[width=\textwidth]{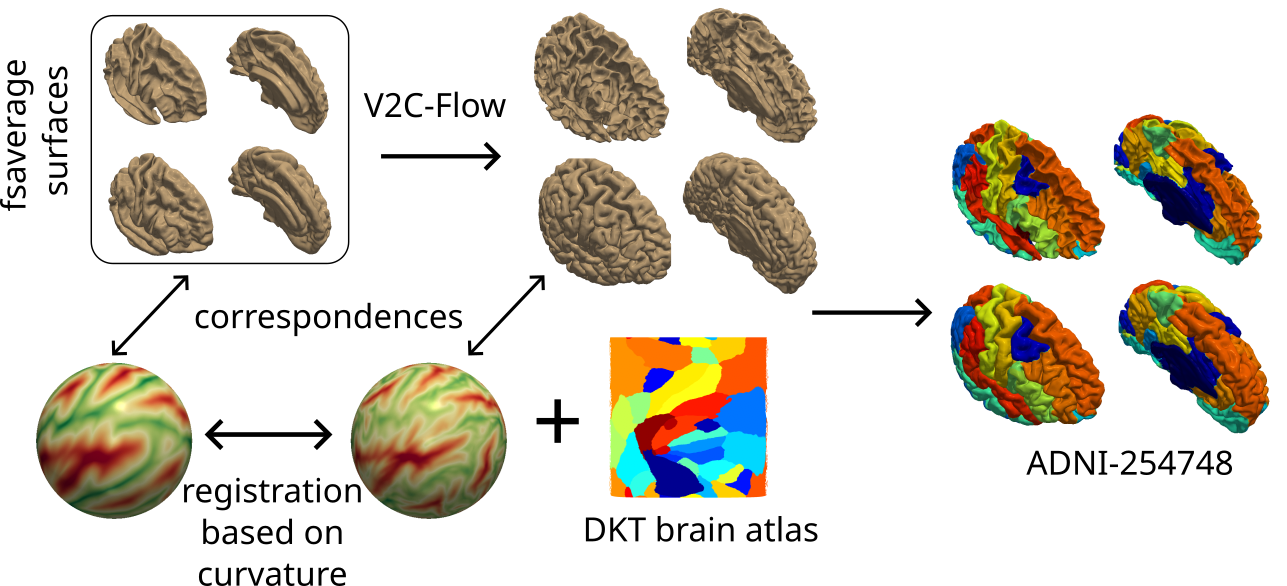}
     \subcaption{}
\end{subfigure}

 \caption{We propose two  approaches to map a parcellation atlas onto V2C-Flow surfaces. (a) The first approach leverages the learned correspondences between the template and individual surfaces. Here, the atlas is mapped directly from the template onto predicted surfaces based on vertex IDs. (2) The second approach exploits the vertex-correspondences to the inflated template sphere. Thereby, the individual curvature pattern can be mapped directly onto the template sphere, which in turn can be used to perform a spherical registration, e.g., using FreeSurfer's \emph{mris\_register}, before assigning vertex classes according to the parcellation atlas. We also show exemplary parcellations obtained with either method for the same sample of our ADNI test set (each color indicates an atlas region).}
    \label{fig:parcellation-concepts}
\end{figure*}

\begin{figure*}
    \centering
\begin{subfigure}[b]{\textwidth}
     \includegraphics[width=\textwidth]{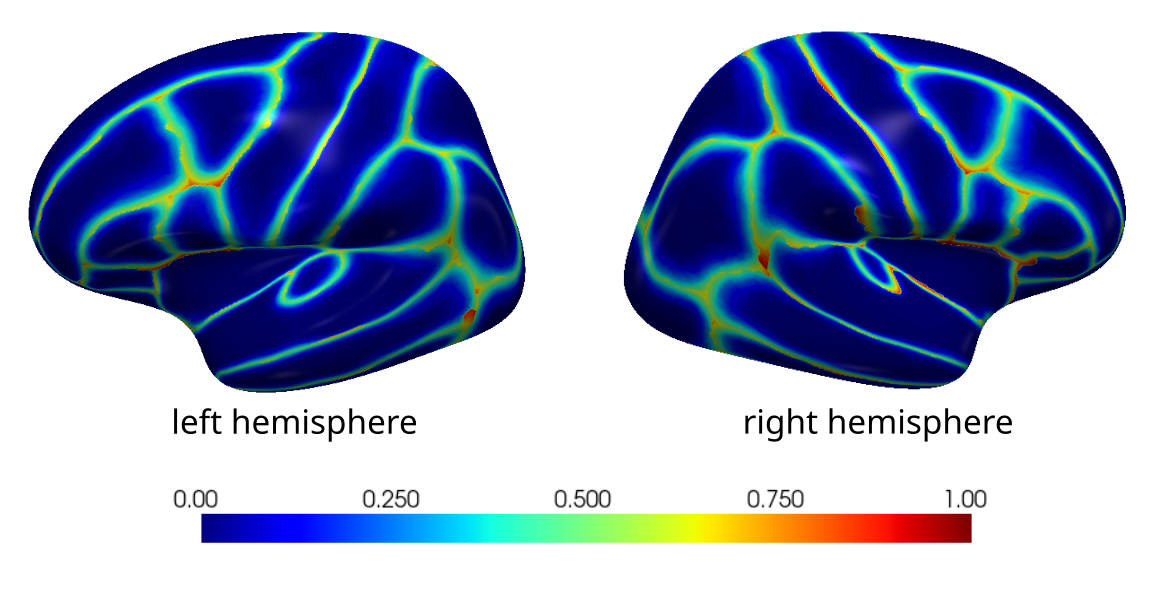}
     \subcaption{}
\end{subfigure}
\begin{subfigure}[b]{\textwidth}
     \includegraphics[width=\textwidth]{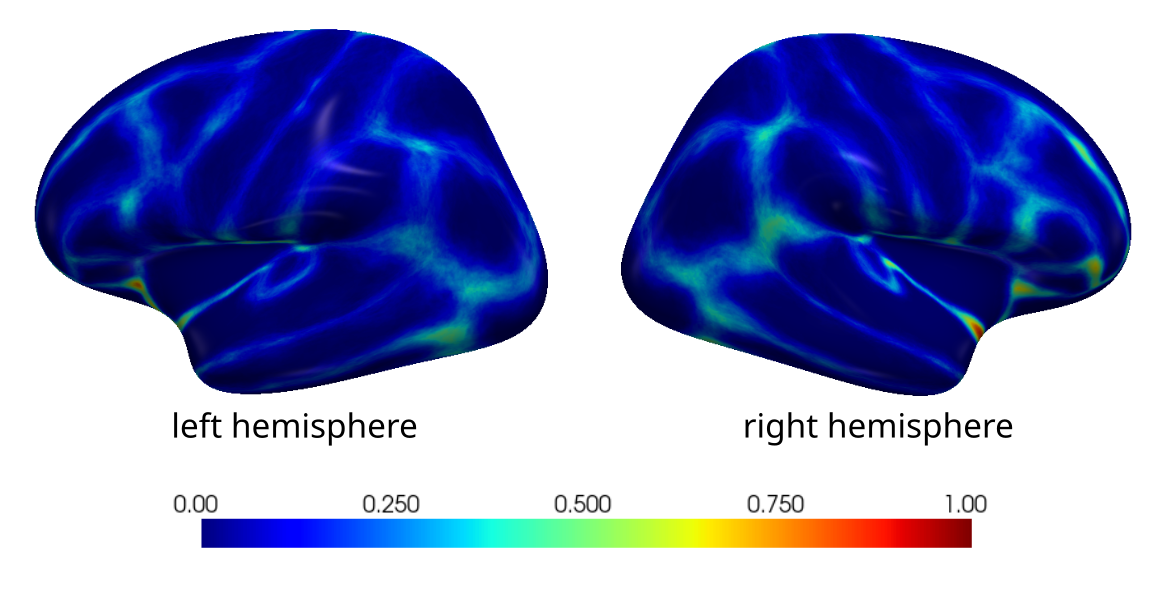}
     \subcaption{}
\end{subfigure}
    \caption{Visualization of parcellation errors for  mapping- (top) and registration-based (bottom) approaches. Errors occur exclusively at parcel boundaries. Colors indicate the frequency of misclassification on the ADNI test set per vertex (0: vertex always classified correctly, 1: vertex always misclassified).}
    \label{fig:parcellation-errors}
\end{figure*}

\begin{figure*}
    \centering
    \includegraphics[width=\textwidth]{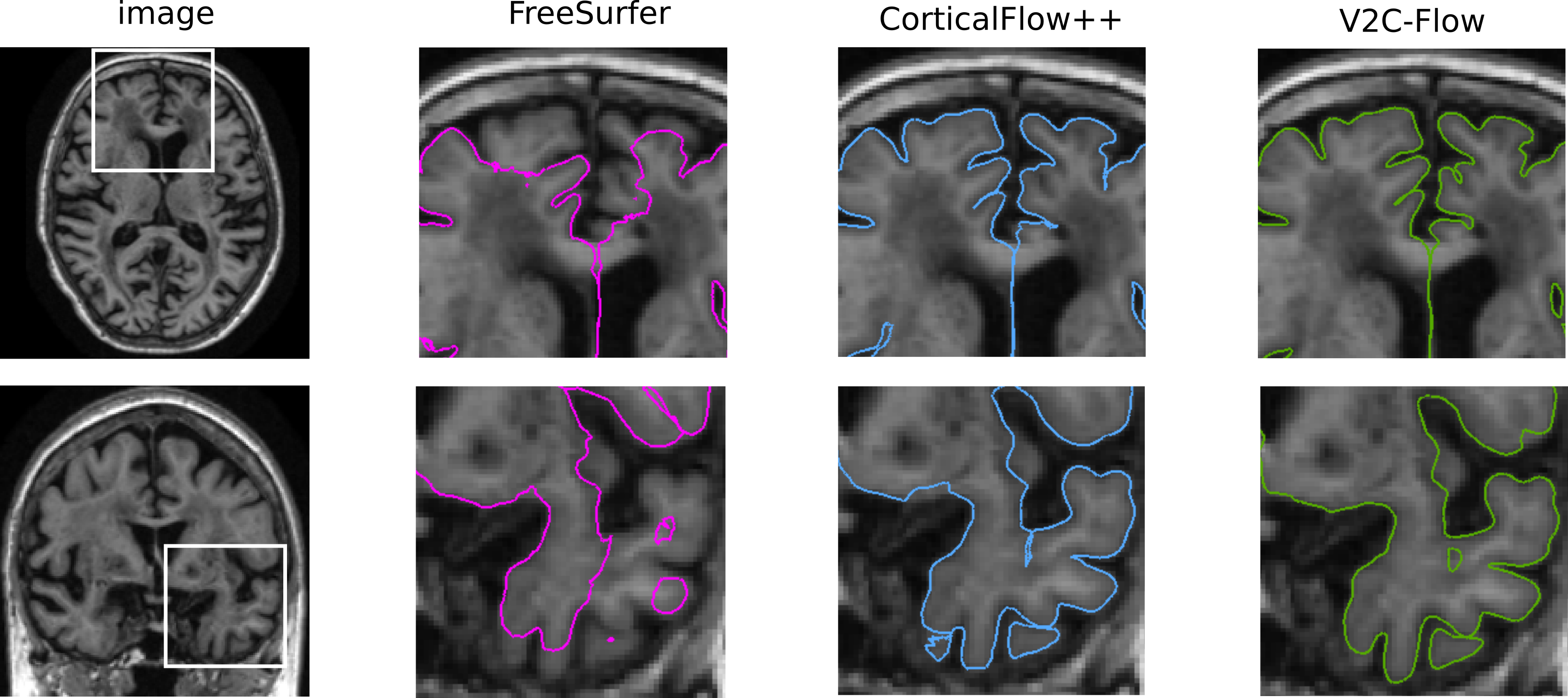}
    \caption{Illustration of an exemplary scan from ADNI with corresponding reconstructed pial surfaces from FreeSurfer, CorticalFlow$^{++}$, and V2C-Flow. We excluded scans with such clear failures of FreeSurfer in our analyses to avoid a negative impact during training or the corruption of test results. Yet, V2C-Flow and CF$^{++}$ still manage to extract meaningful pial surfaces in this case.}
    \label{fig:fs-fail}
\end{figure*}

\begin{figure*}
    \centering
    \includegraphics[width=\textwidth]{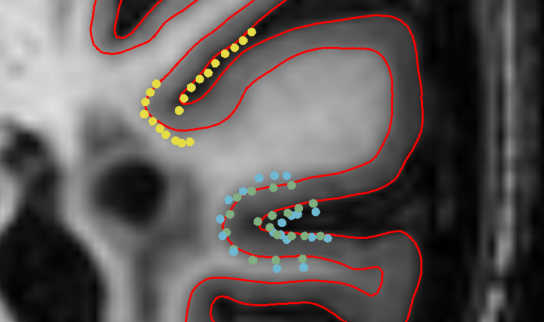}
    \caption{\rev Visualization of manual landmarks (dots) by three raters (indicated by color) and predicted V2C-Flow contours (in red) of WM and pial surfaces next to a white matter lesion.}
    \label{fig:landmarks-lesion}
\end{figure*}

\end{document}